\documentclass[useAMS,usenatbib]{mnras}

\usepackage{graphicx}
\usepackage{verbatim}
\usepackage{bm}
\usepackage{pgf}
\usepackage{mathptmx}
\usepackage{amsmath,amssymb}
\usepackage{adjustbox}
\usepackage{subcaption,import}

\usepackage{ulem}
\usepackage{xspace}
\usepackage{siunitx}
\usepackage{hyperref}
\hypersetup{colorlinks = true}
\usepackage{orcidlink}

\DeclareMathOperator{\var}{var}
\DeclareMathOperator{\err}{err}

\newcommand{\qty}{\SI}
\newcommand{\qtyrange}{\SIrange}
\newcommand{\unit}{\si}

\usepackage{xpatch}
\makeatletter
\xpatchcmd\NAT@citex
 {%
  \@citea\NAT@hyper@{%
    \NAT@nmfmt{\NAT@nm}%
    \hyper@natlinkbreak{\NAT@aysep\NAT@spacechar}{\@citeb\@extra@b@citeb}%
    \NAT@date
  }%
 }
 {%
  \@citea
  \NAT@nmfmt{\NAT@nm}%
  \NAT@aysep\NAT@spacechar
  \NAT@hyper@{\NAT@date}%
 }
 {}{}
\xpatchcmd\NAT@citex
 {%
  \@citea\NAT@hyper@{%
    \NAT@nmfmt{\NAT@nm}%
    \hyper@natlinkbreak{\NAT@spacechar\NAT@@open\if*#1*\else#1\NAT@spacechar\fi}%
    {\@citeb\@extra@b@citeb}%
    \NAT@date
  }%
 }
 {
  \@citea
    \NAT@nmfmt{\NAT@nm}%
    \NAT@spacechar\NAT@@open\if*#1*\else#1\NAT@spacechar\fi
    \NAT@hyper@{\NAT@date}%
 }
 {}{}
\makeatother

\title[Lensing in the eclipse of \psr]{Plasma lensing near the eclipses of the Black Widow pulsar B1957$+$20} 

\author[F. X. Lin et al.]
{F. X. Lin$^{\orcidlink{0000-0002-6820-4275}}$\!$^{1,2}$
\thanks{E-mail:flin@cita.utoronto.ca}, 
R. A. Main$^{\orcidlink{0000-0002-7164-9507}}$\!$^{3}$, 
D. Jow$^{\orcidlink{0000-0003-3236-8769}}$\!$^{1,2}$, 
D. Z. Li$^{\orcidlink{0000-0001-7931-0607}}$\!$^{4}$, 
U.-L. Pen$^{\orcidlink{0000-0003-2155-9578}}$\!$^{1,2,5,6}$,
M. H. van Kerkwijk$^{\orcidlink{0000-0002-5830-8505}}$\!$^{5}$\\
$^{1}$Canadian Institute for Theoretical Astrophysics, 60 St George St. Toronto, Canada \\
$^{2}$Department of of Physics, University of Toronto, 60 St George St. Toronto, Canada \\
$^{3}$Max-Planck-Institut f{\"u}r Radioastronomie, Auf dem H{\"u}gel 69, 53121 Bonn, Germany\\
$^{4}$Cahill Center for Astronomy and Astrophysics, California Institute of Technology, 1216 E California Blvd, Pasadena, CA 91125, US\\
$^{5}$Department of Astronomy and Astrophysics, University of Toronto, 50 St. George Street, Toronto, ON M5S 3H4, Canada\\
$^{6}$Institute of Astronomy and Astrophysics, Academia Sinica,
11F of AS/NTU Astronomy-Mathematics Building, No.1, Sec. 4, Roosevelt Rd,\\ Taipei 10617, Taiwan, R.O.C.\\
}

\newcommand{\msun}{\mbox{M$_{\odot}$}}
\newcommand{\rsun}{\mbox{R$_{\odot}$}}

\newcommand{\psr}{\mbox{PSR B1957$+$20}\xspace}

\newcommand{\veff}{v_\text{eff}}
\newcommand{\vscaled}{v_\text{scaled}}

\newcommand{\dpsr}{d_\text{psr}}

\newcommand{\aorb}{a_\text{orb}}
\newcommand{\vorb}{v_\text{orb}}
\newcommand{\Vorb}{\mathbf{v}_\text{orb}}
\newcommand{\vflow}{v_\text{flow}}
\newcommand{\Vflow}{\mathbf{v}_\text{flow}}
\newcommand{\DM}{\mathrm{DM}}
\newcommand{\ds}{d_\text{s}}
\newcommand{\dl}{d_\text{l}}
\newcommand{\dsl}{d_\text{sl}}
\DeclareMathOperator{\fourier}{\mathcal{F}}

\let\oldhat\hat

\renewcommand{\hat}[1]{\oldhat{\mathbf{#1}}}

\begin{document}

\date{Accepted 2022 November 22. Received 2022 November 8; in original form 2022 August 29}

\pagerange{\pageref{firstpage}--\pageref{lastpage}} 
\pubyear{2022}

\maketitle

\label{firstpage}
  
\begin{abstract}
Recently, several eclipsing millisecond pulsars have been shown to experience strong and apparent weak lensing from the outflow of their ionized companions.
Lensing can be a powerful probe of the ionized plasma, with the strongest lenses potentially resolving emission regions of pulsars. Understanding lensing in the `laboratory-like' conditions of an eclipsing pulsar may be analogously applied to fast radio bursts, many of which reside in dense, magnetized environments.
We examined variable dispersion measure (DM), absorption, scattering, and flux density in the original Black Widow pulsar \psr through an eclipse at the Arecibo Observatory at \qty{327}{MHz}.
We discovered clear evidence of the two regimes of lensing, strong and apparent weak.
We show that the flux density variations in the apparently weak lensing regime can be modeled directly from variations of DM, using geometric optics.
The mean effective velocities in the ingress, \qty{954\pm 99}{km.s^{-1}}, and egress \qty{604\pm 47}{km.s^{-1}} cannot be explained by orbital motions alone, but are consistent with 
significant outflow velocity of material from the companion.
We also show that geometric optics can predict when and where the lensing regime-change between weak and strong occurs, and argue that the apparent weak lensing is due to averaging many images. 
Our framework can be applied in any source with variable electron columns, measuring their relative velocities and distances.
In other eclipsing pulsars, this provides a unique opportunity to measure companion outflow velocity, predict regions of weak and strong lensing, and in principle independently constrain orbital inclinations.
\end{abstract}


\section{Introduction}
\label{sec:intro}

\begin{figure*}
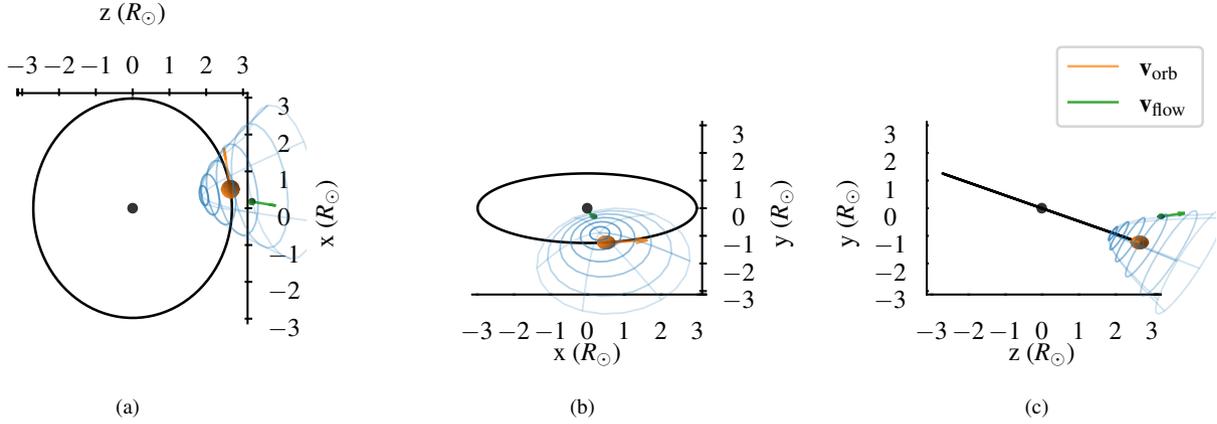

    \begin{subfigure}[t]{.33\textwidth}
      \centering
      \begin{adjustbox}{clip,trim=0.0cm 0.5cm 0.0cm 0cm, max width=0.95\textwidth}
      \input{figures/system_plot_top.pgf}
      \end{adjustbox}
      \caption{}
      \label{fig:sub-first}
    \end{subfigure}
    \hfill
    \begin{subfigure}[t]{.33\textwidth}
      \centering
      \begin{adjustbox}{clip,trim=0.0cm 0.5cm 0.0cm 0cm, max width=0.95\textwidth}
      \input{figures/system_plot_face.pgf}
      \end{adjustbox}
      \caption{}
      \label{fig:sub-second}
    \end{subfigure}
    \hfill
    \begin{subfigure}[t]{.33\textwidth}
      \centering
      \begin{adjustbox}{clip,trim=0.0cm 0.5cm 0.0cm 0cm, max width=0.95\textwidth}
      \input{figures/system_plot_side.pgf}
      \end{adjustbox}
      \caption{}
      \label{fig:sub-third}
    \end{subfigure}
    \caption{Schematics of the \psr binary system, drawn at the orbital phase of $\sim$ 0.28, with the pulsar at the origin. The companion (the orange sphere), the binary separation $\aorb\sin i\approx 2.7 R_{\odot}$ and inclination $\sim 65^\circ$ are drawn to scale. The longitude of ascending node $\Omega$ is assumed to be \ang{0}. The $x$-axis is along the line of nodes, and the positive $z$-axis is towards the observer. In this frame, the pulsar is thought of as being stationary, and the companion is moving with velocity $\Vorb$, shown as the orange arrow. The hyperboloid wireframe is a sketch of the intrabinary shock, without the observed asymmetry. The green sphere shows a blob of outflowing lensing material with velocity $\Vflow$, shown as the green arrow. 
    (a)~Top view. 
    (b)~Front view.
    (c)~Side view.
    }
    \label{fig:system_geometry}
    \end{figure*}

The interstellar medium (ISM) is known to be inhomogeneous. The electron over- and under-densities $\Delta n_e$ in the ISM are proportional to changes in refractive index $\Delta n$. Thus, variations in the electron column density, i.e. dispersion measure (DM) of a pulsar at distance $\dpsr$, defined as
\begin{align}
    \DM\equiv\int_0^{\dpsr} n_e\,dl,
\end{align}
must directly lead to lensing effects. These include multiple imaging, magnification, delays in pulse arrival times, scattering, and many other related effects. However, despite much effort, the link between DM and lensing have been tenuous. One of the most successful examples is the modeling of the 1997 `echo', i.e. a copy a the pulse profile, seen in the Crab pulsar by \citet{backerPlasmaPrismModel2000}. 
The echo of the pulsar, interpreted as a refracted macro-image, preceeded a large increase in DM, likely due to the line of sight crossing a filament of the Crab nebula \citep{grahamsmith1997EventCrab2011}. 
Flux density of quasars during extreme scattering events (ESEs), have also been well modeled from plasma lenses \citep[see, e.g.][]{cleggGaussianPlasmaLens1998}. 
However, since we cannot directly measure DM changes for quasars, which do not emit pulsed emission, the nature of the lenses remains poorly understood.
The inverse problem, determining DM from the light curve, is also interesting, but difficult since the highly non-linear nature of light propagation makes the problem generically non-invertible. 
\citet{tuntsovDynamicSpectralMapping2016} successfully inverted the dynamic spectrum of an ESE in PKS~1939$-$315 to produce a column density profile of the plasma lens.

Interest in understanding and predicting plasma lensing of radio sources has been renewed due to the large population of fast radio bursts (FRBs) which almost all show significant spectral structure in time and frequency. Some FRBs are known to reside in dense, strongly magnetized, and highly variable environments (FRB~110523: \citealp{masuiDenseMagnetizedPlasma2015}; FRB~121102: \citealp{michilliExtremeMagnetoionicEnvironment2018}; FRB~20201124A: \citealp{xuFastRadioBurst2021}; FRB~20190520B: \citealp{anna-thomasHighlyVariableMagnetized2022}).
In these environments, propagation effects are large. While the propagation effects through a magnetized medium are discussed in \citet{liConstrainingMagneticFields2019,liHighlyVariableMagnetized2022,beniaminiFaradayDepolarizationInduced2022}, to better understand FRBs, understanding lensing is important since it is likely to contribute to some of the time and spectral structure \citep[see, e.g.][]{cordesLensingFastRadio2017}. Some of the potential propagation imprints on the spectral structures of bursts from FRB121102, which may be caused by DM structures, were analyzed by \citet{plattsAnalysisTimefrequencyStructure2021}.

A particularly interesting class of pulsars to study for propagation effects are the so-called `spider' (black widow or redback) pulsars. This class of pulsars is often eclipsing, due to the low-mass companions being ablated by emission from the pulsar. The pulsar wind causes the companion to inflate and shed material, which can often be seen as extra electron over-densities near the pulsar superior conjunctions (i.e., eclipses). 
Recently, evidence of both strong and weak plasma lensing, associated with the intrabinary material, has been seen in a few eclipsing spider pulsars near the eclipses \citep{mainPulsarEmissionAmplified2018,bilous+19,linDiscoveryModellingBroadscale2021,wangUnusualEmissionVariations2021}. 
In `weak lensing', the light curve is weakly modulated with root mean square (RMS) modulations $\lesssim 1$, with little chromaticity; in contrast, in `strong lensing', the light curve can be strongly modulated with RMS modulations $\gtrsim 1$ and is typically highly chromatic. 
Conclusive evidence for the causal link between DM and apparent `weak' plasma lensing has been shown by \citet{linDiscoveryModellingBroadscale2021} in the black widow system PSR~J2051--0827.

\subsection{The Black Widow pulsar \psr}
Since its discovery in 1988, the Black Widow pulsar \psr{} has provided a wealth of studies on the evolution history of millisecond pulsars, as well as studies on eclipse mechanisms \citep[see, e.g.][]{fruchter+88,thompson+94,polzin+20}. In addition, depending on the orbital inclination, \psr may be one of the most massive pulsars known with a mass of $\sim \qty{2.4}{\msun}$ \citep{vankerkwijkEvidenceMassiveNeutron2011}. 
\citet{kandelAtmosphericCirculationBlack2020} improved the light curve modeling of \citet{vankerkwijkEvidenceMassiveNeutron2011}, and found a more modest neutron star mass of $\sim \qty{2.2}{\msun}$. 
More recently, evidence of an X-ray eclipse suggests an edge-on inclination of $\sim\ang{90}$ of the system, with neutron star mass of only $\sim \qty{1.8}{\msun}$ \citep{kandelXMMNewtonObservesIntrabinary2021}. 
Lensing is sensitive to the geometry of the system, thus, studying lensing offers an avenue to independently constraint the mass of the pulsar. 

In this paper, we show that similar to PSR~J2051--0827, a `weak' regime exists for \psr, where strong lensing has been previously observed \citep{mainPulsarEmissionAmplified2018}. 
Although the light curve RMS modulations are less than unity, the lensing is only `weak' in the sense that the signal is weakly modulated, akin to `weak scintillation', and the observed scattering implies multiple imaging and hence that the physical lensing is strong. 
Throughout the paper, for brevity, we will use `apparent weak lensing' and `weak lensing' interchangeably to mean the former unless otherwise specified, and return in Section~\ref{sec:weak_manyimg} for discussions on the subject. 
We show that the regime changes from `apparent weak' to strong below a certain critical DM. 
We also argue that geometric optics is applicable to study both regimes, and can adequately explain the transition. 
In addition, we measure the effective velocities of the eclipsing material across the eclipse by fitting DM variations to the measured flux density variations. 

\begin{table*}
  \centering
  \caption{Pulsar and Binary Parameters}
  \label{tab:params}
  \begin{tabular}{lcl} 
    \hline
    Parameter & Value & Note\\
    \hline
    \hline
    DM\dotfill & \qty{29.1098}{pc.cm^{-3}} & Value used for de-dispersion\\
    Spin Period, $P$\dotfill  & \qty{1.607}{ms} & \\
    Orbital Period\dotfill  & \qty{9.2}{hr} &\\
    Projected semi-major axis, $X$ & \qty{0.038}{\rsun} & From timing \citep{fruchter+88}\\
    Projected binary separation, $\aorb\sin i$ & \qty{2.7}{\rsun} & From timing, assuming mass ratio from below \\
    Mass ratio\dotfill & 69.2 & From  \citet{vankerkwijkEvidenceMassiveNeutron2011}\\
    Distance to pulsar, $\ds$\dotfill & $\sim$ \qty{2}{kpc} & Estimated from NE2001 \citep{cordesNE2001NewModel2003} \\
    Relative orbital velocity, $\vorb$& \qty{360}{km.s^{-1}}  & Assuming mass ratio from above \\
    Inclination, $i$\dotfill & 65$^{\circ}$ & From \citet{vankerkwijkEvidenceMassiveNeutron2011}\\
    & $78^{\circ}$ & X-ray light curve modeling if not eclipsed from \citet{kandelXMMNewtonObservesIntrabinary2021}\\
    & $\sim 90^{\circ}$ & If eclipsed by companion photosphere\\
    \hline
  \end{tabular}
\end{table*}

The two most relevant orbital parameters for studying lensing by intra-binary material are the relative orbital velocity
\begin{align}
    \label{eq:relative_velocity}
    \vorb = \frac{K_\text{p}}{\sin i} \frac{m_\text{p}+m_\text{c}}{m_\text{c}},
\end{align}
and the binary separation% is 
\begin{align}
	\aorb = \frac{X}{\sin i }\frac{m_\text{p}+m_\text{c}}{m_\text{c}},
\end{align}
where $K_\text{p}$ and $X$ are the radial-velocity amplitude and projected semi-major axis of the pulsar measured from timing, $i$ the inclination of the system, and $m_\text{p}$ and $m_\text{c}$ are the masses of the pulsar and its companion, respectively.  We outline some of the relevant orbital parameters of \psr{} for our analysis in Table~\ref{tab:params}. We also sketch out the system at the orbital phase $\sim 0.28$ in Figure~\ref{fig:system_geometry} in the pulsar-centered frame, showing the companion, pulsar, and the intrabinary shock.

\subsection{Organization of the paper}
This paper is organized as follows. In Section~\ref{sec:data}, we briefly discuss the observations of the eclipse using the Arecibo telescope and the processing of the data. In Section~\ref{sec:measurements}, we discuss the measurement of variable DM, flux density variations, scattering timescales, and changing lensing regimes. In Section~\ref{sec:model}, we briefly introduce the lensing formalism, and discuss our fit of DM power spectra to flux density variation power spectra, and show that there is strong positive correlation during the entirety of the weak lensing regime in the egress. 
In Section~\ref{sec:velocity}, we discuss the possible reasons for measuring a larger effective velocity for the eclipsing material than the relative orbital velocity,
and show that a significant outflow velocity from the companion could explain the observed velocities.
In Section~\ref{sec:interpretation}, we summarize and discuss the lensing phenomenon as a whole in \psr. Finally, in Section~\ref{sec:summary}, we discuss some potential venues to apply our method.

\section{Observations}
\label{sec:data}
We recorded baseband data for five different eclipses of \psr at the Arecibo Observatory using the \textit{William E. Gordon} (\textit{Arecibo}) Telescope from April 2018 to July 2018 at P-band and \qty{430}{MHz} using the PUPPI backend. 
For this work, we chose to analyse the P-band data recorded on MJD 58261 (23 May 2018) for \qty{1}{hr} \qty{45}{min} with 32 frequency channels over \qtyrange{277}{377}{MHz} which shows the full range of effects under study. Because of the bandpass filter, only the middle $\sim\qty{60}{MHz}$ frequency is usable. Otherwise, we retain the 32 frequency channels from PUPPI.
We dedispersed the recording to a DM of \qty{29.1098}{pc.cm^{-3}}, and use 128 phase gates for each pulsar rotation. 
For the rest the paper, for brevity, we will use `DM' to refer to the excess DM due to the intra-binary material relative to this de-dispersed value. 

\section{Measurements}
\label{sec:measurements}
\begin{figure*}
    \begin{center}
        \begin{adjustbox}{clip,trim=0.5cm 0cm 0.3cm .3cm, max width=0.95\textwidth}
        \input{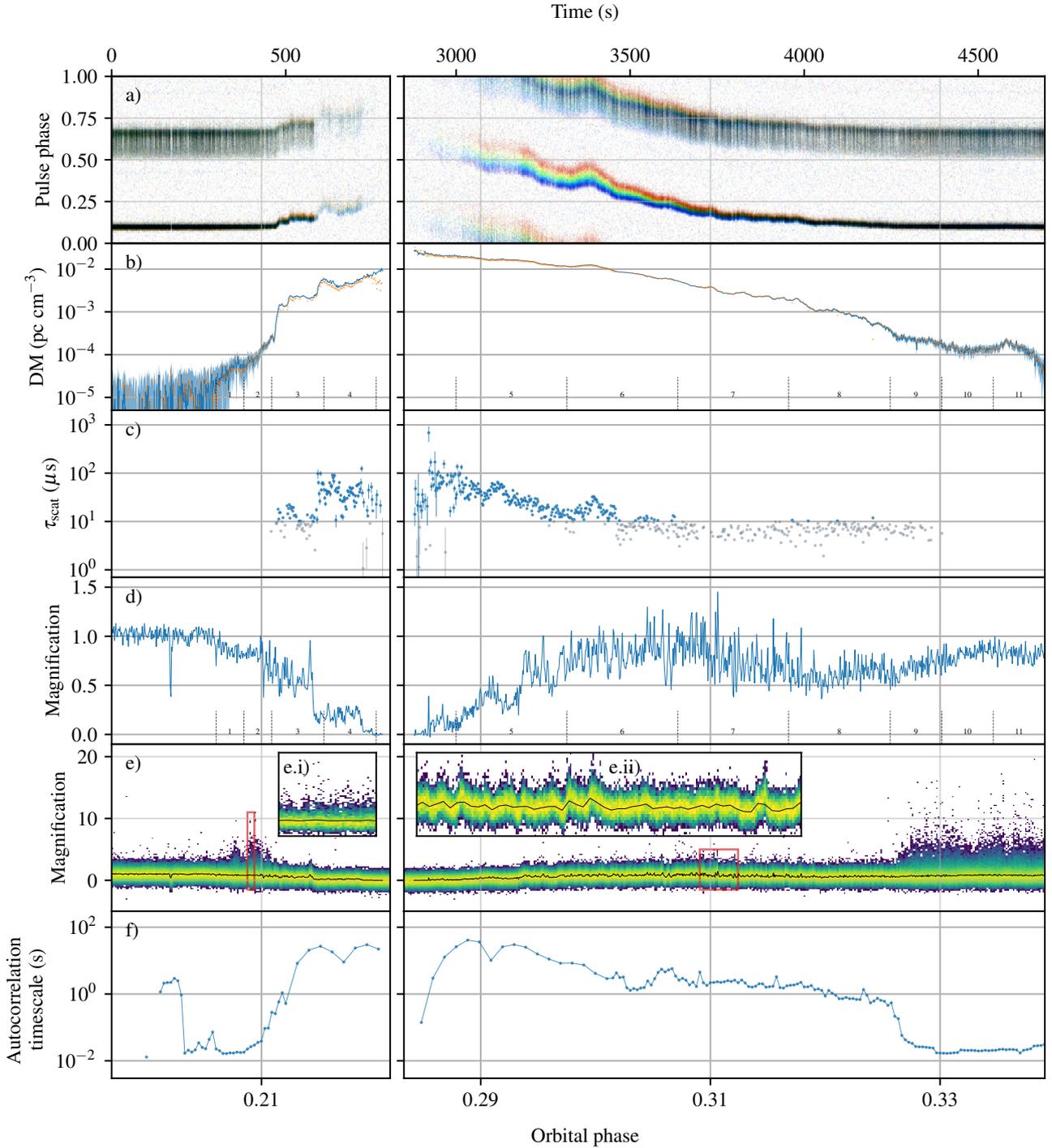}
        \end{adjustbox}
    \end{center}
    \caption{
    Measurements of \psr through an eclipse from the observation on MJD 58261. 
    \textbf{a)} Pulse profile through the eclipse, which becomes visibly dispersed and scattered as the pulsar approaches eclipse. The red through blue are the low through high frequencies of our observing band, respectively. The `flicker' in the interpulse is due to mode-switching. 
    \textbf{b)} DM through the eclipse. Blue line marks the DM measured in 50-pulse folds, assuming no scattering (rebinned to $\sim\qty{1.5}{s}$ for clarity). The blue shade indicates measurement error, with an additive factor of \qty{2.5e-5}{pc.cm^{-3}} to account for mode-switching as discussed in Section~\ref{sec:measurements}. Orange line shows the DM measured in 2-s folds where we also simultaneously measure scattering (see Section~\ref{sec:measurements}). Note that the general trend is very similar, but the blue curve is larger where scattering is significant. For the orange curve, only points with scattering outside the range \qtyrange{0}{10}{\us} exclusive are plotted, since inside this range the measurement is unreliable due to the DM and scattering having similar effects on the pulse profile (see Section~\ref{sec:measurements}). The dashed dividers indicate the regions that power spectra in Figure~\ref{fig:magdmPS} are taken from. 
    \textbf{c)} Scattering timescale at our central frequency \qty{327}{MHz}. Blue points are the measured scattering times $\geq \qty{10}{\us}$, which we deem to be reliably measured. Grey points are the scattering times $<\qty{10}{\us}$, which are unreliable (see Section~\ref{sec:measurements}).
    \textbf{d)} Frequency-averaged, scatter-corrected magnification in \qty{2}{s} folds. 
    \textbf{e)} 2D histogram of single-pulse magnifications (not scatter-corrected) in 5-s bins. The colorbar is in log counts, and transparent pixel means 0 pulses in bin has this particular magnification. The black line is the same curve as in d).
    \textbf{e) i--ii)} Zoom on the rectangle regions. Histograms in the zoom are binned to \qty{0.5}{s}
    \textbf{f)} Autocorrelation timescale of the single-pulse magnifications. Two separate resolutions are used. From \qtyrange{0}{500}{s} and \qty{3450}{s} onwards, we use a 60-s folding timescale. For \qtyrange{500}{3450}{s}, we use a 200-s folding timescale to account for the $\sim$ minute long modulations, which can be seen in panel d). 
    }
    \label{fig:measurements_58261}
\end{figure*}
As the pulsar moves behind its companion, pulses get visibly dispersed and scattered (see panel a) of Figure~\ref{fig:measurements_58261}).
The pulsar also becomes fainter, likely due to absorption by materials in the companion wind. 
Mode-switching of \psr is clearly visible by the flickering brightness of the interpulse. 
The main pulse, whose mean intensity does not vary significantly between the two modes \citep{mahajanModeChangingGiant2018}, also lightly fluctuates in brightness near the eclipse due to lensing by intra-binary material, as described in Section~\ref{sec:model}. 
To simultaneously measure excess DM, scattering times, and flux density, we use the phase-frequency domain template matching method employed in \citet{linDiscoveryModellingBroadscale2021} based on \citet{pennucciElementaryWidebandTiming2014}'s method. We refer the readers to these papers for full details.

Simultaneously measuring DM and scattering time $\tau$ is difficult for the
folding timescales smaller than \qty{2}{s} using a template matching method since decreasing DM and increasing $\tau$ can give a very similar profile as the original profile. To get finer time resolution DM measurements, we take the scattering time to be 0, which allows DM measurements on the folding timescales of 50 spin periods (\qty{80}{ms}), at the cost of over-estimating DM where scattering is significant. We therefore have two sets of DM (and flux density) measurements: one which is scatter corrected, at 2-s folding; and one which is not scatter corrected, at 80-ms folding. 

Panel b) in Figure~\ref{fig:measurements_58261} shows both non-scattered corrected DM as the blue line, rebinned by averaging on \qty{1.5}{s}, and the scatter-corrected DM in \qty{2}{s} bins as orange points. For the scatter-corrected DM, only points with scattering $\geq\qty{10}{\us}$ or $=\qty{0}{\us}$ (bounded by the fitting routine) are shown. 
Panel c) shows the scattering timescale at the central frequency, 327 MHz, and panel d) shows the scatter-corrected pulse amplitude in \qty{2}{s} resolution from the template matching method. Since a \qty{10}{\us} scattering timescale corresponds to $\sim$ 0.5 pixel shift between the top and bottom of the band, the simultaneous measurement of DM and scattering is likely unreliable for scattering timescale of less than \qty{10}{\us}, which are the points shown in grey. 
We use the uncorrected DM time-series for our modeling. The difference between the two time-series is $\sim 0$ when there's no measurable scatter, $\sim\qty{2.5e-4}{pc.cm^{-3}}$ when $\tau\simeq\qty{20}{\us}$, and going up to $\sim\qty{1.5e-3}{pc.cm^{-3}}$ closer to the eclipse where $\tau\gtrsim\qty{60}{\us}$. 

A different challenge exists for profile matching to measuring the 50 pulse fold resolution DM curve. 
\psr exhibits significant mode-switching on $\sim$ seconds timescale \citep{mahajanModeChangingGiant2018}. 
The interpulse flux differs significantly between the `low-' and `high-' modes (see, e.g. panel a) of Figure~\ref{fig:measurements_58261}), but the main pulse is relatively stable. 
We therefore first create a `mode lookup table' by folding the pulsar to 100 rotations, sufficient to distinguish between the two modes.
We then compute the relative ratio of flux between interpulse and main pulse, comparing it against a threshold, and assume that the mode stays constant within each 100 pulses, an order smaller than the average time between mode transitions $\sim 1000$ pulses \citep{mahajanModeChangingGiant2018}. 
For each point of the DM curve, we use the template of the respective mode from the lookup table to fit the DM. 
This introduces systematic noise if we fit the `wrong' mode, for example, if the pulsar is transitioning between two modes within a bin. 
We find the upper bound on the bias in selecting the wrong mode by measuring the DM of 1000 simulated pulses dispersed to a fiducial DM of \qty{1e-3}{pc.cm^{-3}} using low-mode profiles then fitting with high-mode, and vice-versa, with signal-to-noise ratio roughly equivalent to 50 folded pulses.
In each case, the bias is \qty{2.5e-5}{pc.cm^{-3}}, and can be attributed to the minor shift in the position of the centroid of the main pulse in the two modes.
To capture this in our error estimate, we add the bias to our measurement error $\sigma_\DM$ in quadrature.

To measure the pulse-to-pulse magnification, we first de-disperse the pulses using the measured DM timeseries. 
Due to the prominent mode-switching, we only use the main pulse to define pulse-to-pulse magnification.
The magnification of a pulse, measured in an 8 phase gate window ($\sim$ \qty{100}{\us}) around the main pulse, is then the ratio of its flux to the mean flux in a region far away from the eclipse. 
We note that this tends to under-estimate the magnification in the highly scattered region around the eclipse since some of the the main pulse flux may be smearing beyond our phase gate (each gate is \qty{12.5}{\us}). %
Panel e) of Figure~\ref{fig:measurements_58261} shows the 2D histogram of single pulse magnifications in \qty{5}{s} windows. 
Panel f) shows the auto-correlation timescale of the pulse-to-pulse variations. The change of lensing regime can be seen accompanied by a sharp jump in timescale. The second long timescale near an orbital phase of 0.205 is likely due to the `mini-eclipse' seen at the beginning of the data.

\begin{figure*}
    \begin{center}
        \begin{adjustbox}{clip,trim=0.0cm 0cm 0.0cm 0cm, max width=0.9\textwidth}
        \input{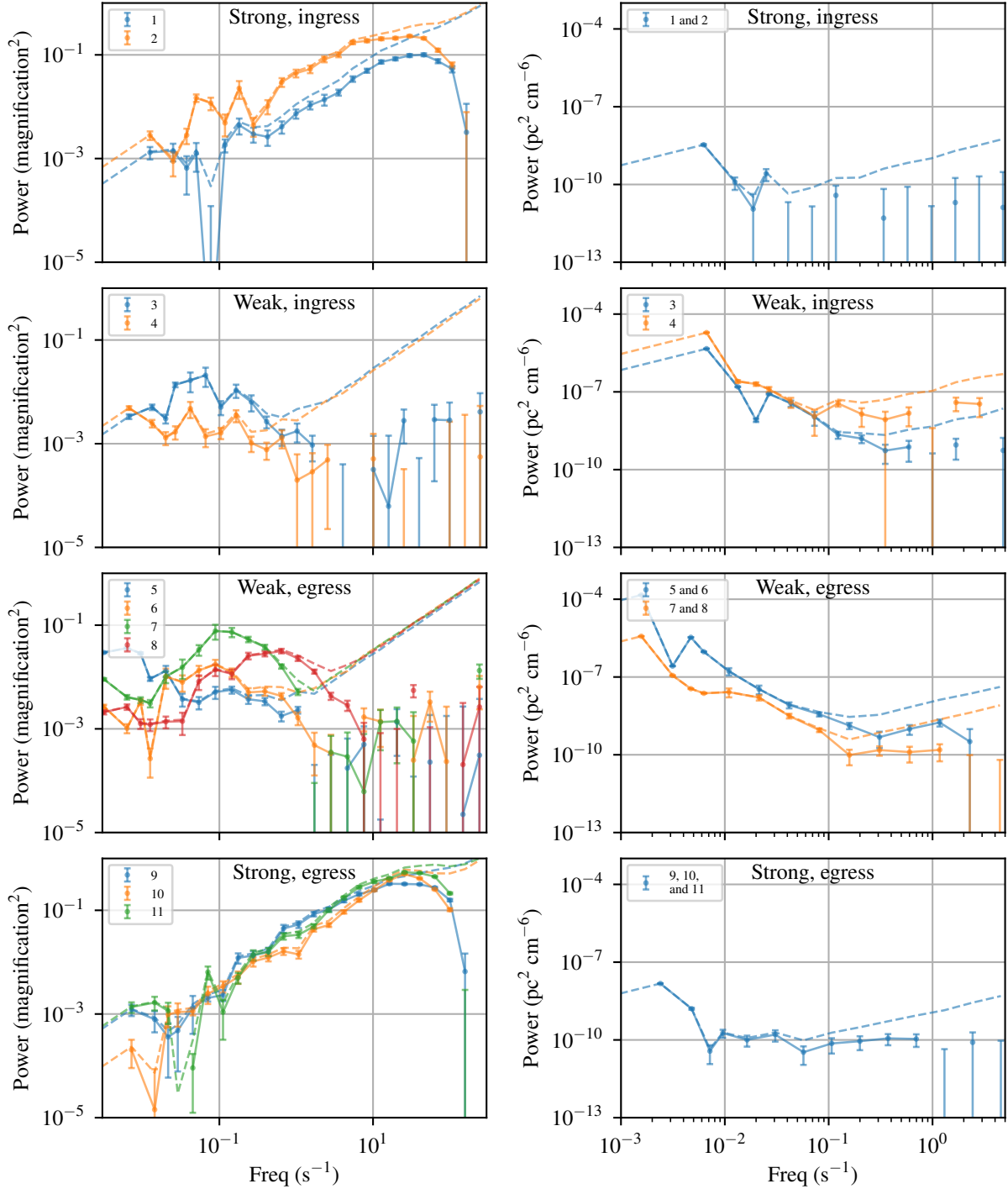}
        \end{adjustbox}
    \end{center}
    \vspace{-5mm}
    \caption{
    Comparing the power spectra of magnifications $\Delta_\mu^2$ (left) and DM $\Delta_\DM^2$ (right) for strong and weak lensing at different sections of the data on MJD 58261 with (solid line) and without (dashed line) a constant noise bias subtracted, as indicated in the labels sections in panels b) and d) of Figure~\ref{fig:measurements_58261}. The different coloured lines indicate the different sections used.
    Error bars are discussed in Section~\ref{sec:lensing_regimes}.
    The location where the dashed lines begin to `turn over' shows the timescales below which noise begins to dominate (e.g. $\lesssim$ few seconds for weak lensing magnification spectra and DM spectra). 
    In the magnification spectra (left), in the weak lensing regime, power peaks at low frequencies, i.e., $\gtrsim$ seconds timescales. 
    The DM power spectra (right), when measurable, i.e. in the middle two panels, follow a power-law index of $\sim -3$ (which looks like $-2$ in this space, due to the multiplication by $f$ in Equation~\ref{eq:dimless_powerspec} ).
    Our model fits the DM power spectra to the magnification spectra (with different sections and binnings, see Section~\ref{sec:implementation} for details). 
    On the other hand, the strong lensing region has power that extends to high frequencies (i.e. $\sim$ few-ms timescales) and peaks there.
    }
    \label{fig:magdmPS}
\end{figure*}

\subsection{Lensing regimes}
\label{sec:lensing_regimes}
From Figure~\ref{fig:measurements_58261}, the lensing modulations are starkly divided into two different regimes: a `weak' regime with moderate modulations in brightness closer to the eclipse in both the ingress and egress, with a $\sim$ second to minute timescale, and a `strong' regime farther from the eclipse, with timescales of milliseconds. The transition between the two occurs very fast relative to the duration of the eclipse, which can be seen in, e.g. the 2D histogram of magnifications in e) of Figure~\ref{fig:measurements_58261}, as well as the autocorrelation timescale in panel f) which jumps near the transition. We describe the two regimes in this section. 

In addition to the autocorrelation timescales, the lensing modulation at different timescales can also be quantified using the `dimensionless' power spectra of magnification. 
The left column of Figure~\ref{fig:magdmPS} shows the `dimensionless' magnification power spectra in the ingress and egress for the weak and strong regimes, with different colors describing different subsections of the data as indicated in the caption. 
The `dimensionless' power spectrum\footnote{Not actually dimensionless here since we don't normalize by the mean, but has units of the quantity measured squared.} $\Delta^2_g (f)$, a quantity which measures the contribution to fluctuations per logarithmic frequency or wavenumber interval, typically used in cosmology to describe fluctuations of overdensities of $g(t)$, is defined as
\begin{align}
    \label{eq:dimless_powerspec}
    \Delta^2_g (f) \equiv f\,P_g (f),
\end{align}
where $P_g(f)=|\fourier[g](f)|^2/T$ is the power spectrum (or power spectral density) of $g(t)$, $f$ is the conjugate frequency, $T$ the total time, and we adopt the Fourier transform convention that $\fourier[g](f)\equiv\int dt \exp(-2\pi i ft) g(t)$. 
To avoid confusion, we will simply refer $\Delta^2_g$ as a power spectrum for the rest of the paper. 
For the unbinned frequencies, the error bars for the power spectrum can be obtained from a relatively straightforward calculation (see e.g. \citet{hoyngErrorAnalysisPower1976}). For the binned frequencies, the errors are simply the standard error in the mean of the bins.
Note that in this representation, white noise is not constant, but will grow as $f$. A constant (`noise bias') may be subtracted from $P_g (f)$ to get a better estimate of the intrinsic $\Delta^2_g (f)$. For gaussian white noise,  the bias can be estimated by averaging high enough frequency modes, where the noise floor is apparent. 

\subsubsection{Apparently `weakly' lensed and scattered}
\label{sec:weak_scattering}
The apparent weak lensing region is characterized by its moderate, achromatic modulations in magnification and relatively long lensing timescales. Magnification in this region does not exceed $\sim 2$.
As shown in the left column of Figure~\ref{fig:magdmPS}, in addition to the interstellar scintillation peak on the $\sim$ minutes timescale, weak lensing peaks near $\sim$ 10 s both in the ingress and egress. At an orbital velocity of $\sim\qty{360}{km.s^{-1}}$, this corresponds to a lens of roughly $\sim\qty{3600}{km}$. We show the ingress weak lensing region divided into two sections (blue and orange), and the egress regions divided into four sections. The peaks shifts from a few tens of seconds near the eclipse, to a few seconds near the transition to strong lensing. Additionally, scattering only measurably occurs in the weak lensing region, close to the eclipse. 

The scattering timescale at our middle frequency \qty{327}{MHz} is shown in the panel c) of Figure~\ref{fig:measurements_58261}, rising from the just measurable \qty{20}{\us} at an orbital phase of $\sim 0.30$ to up to \qty{150}{\us} very close to the eclipse around the egress, corresponding to a physical size of scattering disks of $\sim\qty{4800}{km}$ to $\sim\qty{13000}{km}$, or a smoothing scale of $\sim\qty{13}{s}$ to $\sim\qty{35}{s}$ in the measured data. This will the limit our measurement of the DM at finer scales closer to the eclipse, and may explain some of the (lack of) correlation seen in Section~\ref{sec:correlation} between orbital phases of 0.285--0.30. 

Independently, \citet{baiDetectionStrongScattering2022} measured scattering of \psr{} at  \qty{1250}{MHz} surrounding the eclipse.
At \qty{1250}{MHz}, the eclipse is much shorter in duration, spanning orbital phase of $\sim$ 0.241--0.271.
As such, \citet{baiDetectionStrongScattering2022} probe. the system much closer to superior conjunction.
Additionally, significant flux density variations on roughly 30-s long timescale can be seen near the eclipse at \qty{1250}{MHz}, suggestive of a similar weak lensing mechanism at work at higher frequencies.

\subsubsection{Strongly lensed}
\begin{figure*}
    \begin{center}
        \begin{adjustbox}{clip,trim=0.0cm 0cm 0.0cm 0cm, max width=.95\textwidth}
        \input{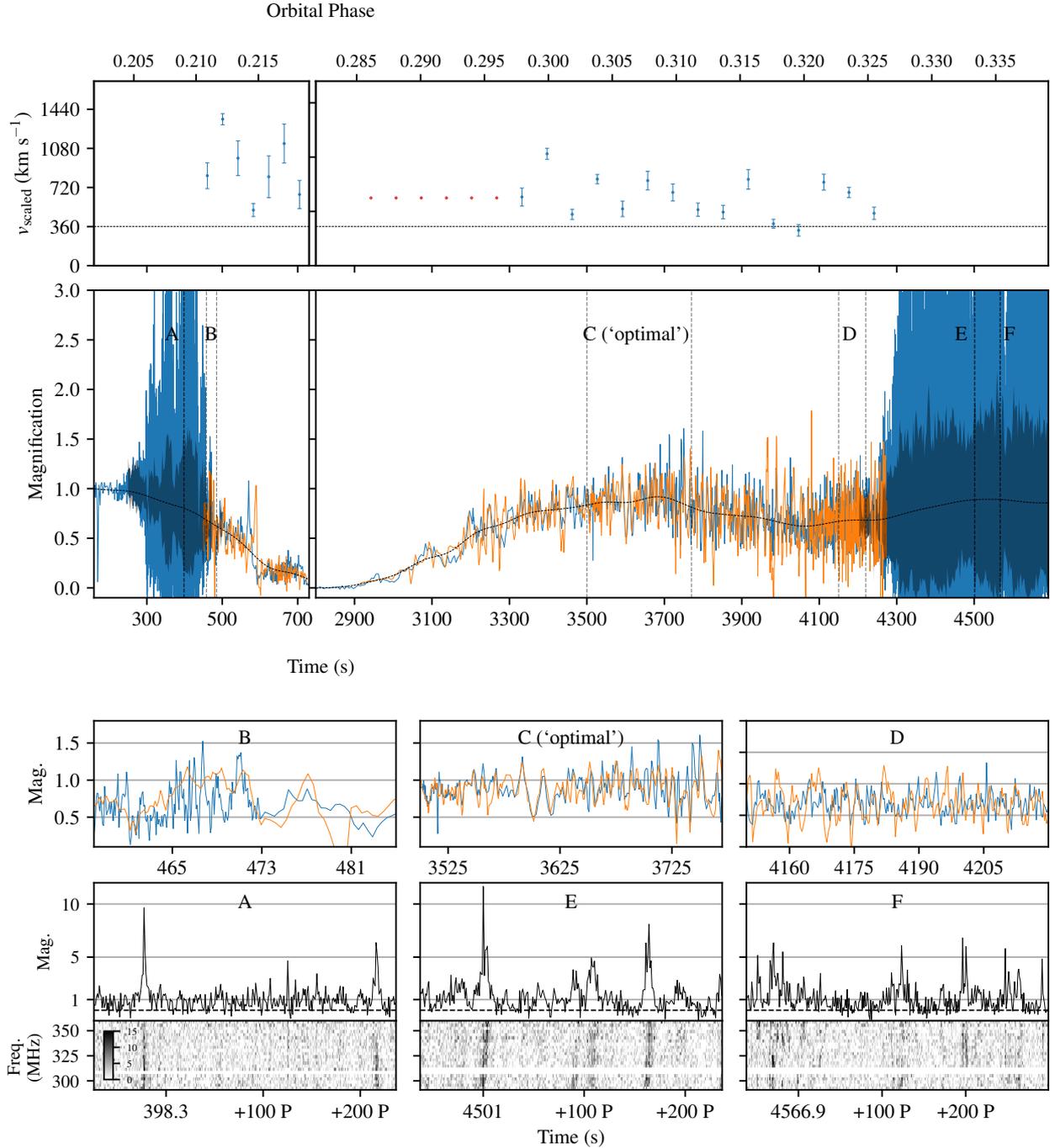}
        \end{adjustbox}
    \end{center}
    \vspace{-5mm}
    \caption{
    Magnification fits from measured DM. The top panels shows the velocity fits (blue points) in sections of \qty{40}{s} in the ingress and \qty{65}{s} in the egress.
    The red sections in the egress are assumed to be at the best fit mean egress velocity \qty{604}{km.s^{-1}}. The black dashed line is the relative orbital velocity \qty{360}{km.s^{-1}}.
    The middle panels shows the measured (blue) and model (orange) light curves, adaptively rebinned by averaging using measured lensing timescales to reduce visual clutter. The dark blue-grey shaded region in the strong lensing regions show the reduced standard deviation $\sqrt{\var(\mu_\text{strong}) - \var(\mu_\text{weak}})$, where $\var(\mu_\text{strong})$ is the variance inside the strong lensing region and $\var(\mu_\text{weak})$ is the variance in the weak lensing region to give a better sense of the magnification fluctuations. The black dashed line is the background model $\mu_\text{bg}$. The model provides a good fit in the time domain from $t\gtrsim\qty{500}{s}$ to $t\lesssim\qty{700}{s}$ in the ingress and a very good fit from $t\gtrsim\qty{3200}{s}$ to $t\lesssim\qty{4000}{s}$. The sections A, B, C, D, E, and F are zoomed in and shown on the bottom panels.
    Sections A, E, and F show a few of the chromatic, strong lensing events first shown in \citet{mainPulsarEmissionAmplified2018} on the bottom, and the frequency-averaged magnification on top. The $x$-axis shows the time in seconds and $P=\qty{1.607}{ms}$ is a single pulse period. Note that close to the bright events, magnification is closer to 0 (black, dashed line) than 1 (clearly seen in e.g. the right-most event in panel E). These are essentially `shadows' on either side of the bright peak. Section B and D show the weak lensing regions with fairly fast lensing timescales, where the time domain fit is only statistically correlated. Section C shows a subsection of the weak lensing region whose timescale is in the `optimal' zone, which provides remarkably good fit in the time domain. 
    }
    \label{fig:fit}
\end{figure*}

The strong regime, markedly different from the weak by its large, chromatic modulations and very short timescales, was first reported by \citet{mainPulsarEmissionAmplified2018}. 
We show examples of such lensed events in the observation in Figure~\ref{fig:fit}, section A, E, and F, and discuss a different interpretation of the events in Section~\ref{sec:geomopt_interpretation}. 
Magnification, averaged over our entire \qty{50}{MHz} band, can reach $\sim 30$. As in \citet{mainPulsarEmissionAmplified2018}, the magnification `events' last a few pulses, i.e. $\sim\qty{10}{ms}$. This can also be seen in the top and bottom panels in the left column of Figure~\ref{fig:magdmPS}. Both in the ingress and egress the strong lensing regime peaks at a few tens of milliseconds, since the power spectrum picks up on the duration of the `shadows' between bright lensing events.  

\section{Modeling weak lensing}

\label{sec:model}
In the weak lensing region, we can model the flux density variations using the measured changes of DM as in \citet{linDiscoveryModellingBroadscale2021}. From the middle two rows of Figure~\ref{fig:magdmPS}, only a subset of the weak lensing region peaks at longer scales than the scale at which DM becomes noise-dominated, around above $\gtrsim\qty{10}{s}$.
In the regions where the magnification power spectrum is dominated by scales below \qty{10}{s}, DM is noise-dominated, thus any model of magnification from DM will be highly dependent on the de-noising method, although a statistical correlation may still exist.

The best fitted lensing model from DM to flux density variations in the `eclipses' of the black widow pulsar PSR~J2051$-$0827 in \citet{linDiscoveryModellingBroadscale2021} appears to be solidly in the linear regime, i.e. single-image, RMS modulations below unity, and with small deflection angles. Instead of fully solving the lens equations, we take inspiration and proceed using linear methods, differing from previous work. 
An interesting `spike' feature exists in the light curve in the ingress, near the orbital phase of $\sim 0.215$, with a magnification of $\sim 2$, and therefore may not be amenable to linear methods. We analyze the `ingress spike' in Section~\ref{sec:ingress_spike}. 

\subsection{Theory}
To start, consider the 2D lens equation
\begin{align}
\label{eq:lens_eq}
    \mathbf{x}_s = \mathbf{x} - \frac{\lambda r_e R_F^2}{2\pi} \nabla_\mathbf{x}\,N_e(\mathbf{x}),
\end{align}
where $\mathbf{x}_s$ is the source position, $\mathbf{x}$ the position on the lens plane, $r_e$ the classical electron radius, $R_F = \sqrt{\lambda \dl \dsl/\ds}\simeq\sqrt{\lambda \dsl}$ the Fresnel scale (with $\dl$, $\ds, $ $\dsl$ the observer-lens, observer-source, source-lens distances, respectively), and $N_e(\mathbf{x})$ the column density of electrons, and the equation for magnification (essentially, the differential ratio of stretched image size $\Delta\mathbf{x}$ to the source size $\Delta\mathbf{x}_s$)
\begin{align}
\label{eq:mag_eq_2d}
    \mu &= \det\left(\frac{\partial\mathbf{x}}{\partial\mathbf{x}_s}\right).
\end{align}

We immediately make several simplifying assumptions. First, we can only measure observables in a single dimension: along the motion of the pulsar. We therefore use the 1D analog of the 2D equations. Second, the physical quantity $N_e$ is never actually observable, and the only observable quantity is the DM, which is \textit{a priori} influenced by lensing at the observer. We assume that the bending of light is not too large, and take $N_e\simeq \DM$ \citep[see][for discussions]{linDiscoveryModellingBroadscale2021}.

In this case, the magnification becomes
\begin{align}
\label{eq:mag_eq}
    \mu_{\DM}&=\left(1-\frac{\lambda r_e R_F^2}{2\pi}\partial_x^2\,\DM(x)\right)^{-1}\nonumber\\
    &\simeq 1+\frac{\lambda r_e R_F^2}{2\pi}\partial_x^2\,\DM(x),
\end{align}
where we take the first two terms of the Taylor expansion in the second line and take $\mu_{\DM}$ to mean \textit{model} magnification. 
The expansion assumes a coherent lens with a single image being magnified (but see \ref{sec:weak_manyimg} for discussions).
From Equation~\ref{eq:mag_eq}, one can in principle take the measured DM, compute two derivatives, and compare to the measured magnification. This is difficult in the presence of noise: each derivative enhances the noise, and thus the results need to be filtered. We opt for a Wiener filter, but note that our results do not depend strongly on the filters that we tested.

The most important parameter of the fit is the conversion factor between the spatial derivatives of $\DM(x)$ appearing in the lensing equations, and the time that $\DM(t)$ is measured in. The conversion factor, an effective velocity, can be shown to be 
\begin{align}
\label{eq:effective_velocity}
    \veff = (\Vorb-\mathbf{v}_{\text{flow}})\cdot\bm{\ell}\,\cos\alpha,
\end{align}
where $\bm{\ell}$ is a unit vector in the direction of motion projected onto the plane of the sky, $\mathbf{v}_{\text{flow}}$ is the flow velocity of the lensing material (see Figure~\ref{fig:system_geometry}), and $\alpha$ is an angle between the lensing axis and the direction of motion since we are using a 1D model \citep{linDiscoveryModellingBroadscale2021}. 
This is the same effective velocity as the effective velocity in literature describing interstellar scintillation arcs \citep[e.g.][]{walkerInterpretationParabolicArcs2004,cordesTheoryParabolicArcs2006,brisken+10}.

For convenience, we define a scaled velocity
\begin{align}
    \label{eq:scaled_velocity}
    \vscaled &= \sqrt{\frac{\aorb\sin i}{\dsl}} \veff \nonumber\\
    &\equiv u^{-\frac{1}{2}} \veff,
\end{align}
where $\aorb \sin i$ is the projected binary separation, $\dsl\equiv u\cdot \aorb\sin i$ is the source-lens distance.

To summarize, we will model the observed magnification variations in the weakly lensed section using the measured excess DM variations and the magnification equation~(\ref{eq:mag_eq}). This involves fitting for a velocity $\vscaled$, related to the geometry of the system and velocity of the lensing material. The fit is performed in Fourier domain. We discuss the details of the fitting routine and best fit velocities in the following section.

\subsection{Implementation}
\label{sec:implementation}
\subsubsection{Practical considerations}
Two additional complications in the lensing fit arise due to the influence of absorption and the influence of interstellar scintillation (ISS), both of which are visible in panel d) in Figure~\ref{fig:measurements_58261}. We assume that the decrease in flux density in the eclipse is due to absorption which acts smoothly on the light curve. ISS acts on $\sim$ tens of \si{kHz} frequency scales, and has power on $\sim$ minutes timescales \citep{mainDescatteringGiantPulses2017}, modulating the light curve by about 10\%--20\%, as can be seen in panel d) of Figure~\ref{fig:measurements_58261} after orbital phase 0.31. 

To account for these effects, we Gaussian smooth the measured pulse-to-pulse magnification with a width of \qty{1}{min}, and use that as a `background' light curve change $\mu_\text{bg}$, which acts multiplicatively with the magnifications from the model. This is a reasonable assumption as long as the variations in flux density due to lensing by intrabinary materials are fast. There is some freedom in choosing the width of the smoothing filter: we have chosen it so that it is longer than the longest intrabinary lensing timescales, close to the eclipse. The model light curve is then
\begin{align}
    \label{eq:model_mag}
    \mu_\text{model}(t) =&\, \mu_\text{bg}(t) \mu_\DM(t) \nonumber\\
    =&\,\mu_\text{bg}(t)\nonumber\\
    &\left(1+\fourier^{-1}\left[(2\pi i f)^2 \left(\frac{\lambda^2 r_e\,\aorb\sin i}{2\pi \vscaled^2}\right) W(f) \fourier[\DM(t)] \right]\right)
\end{align}
for a given $\vscaled$. The second term in the second line comes from the Fourier transform of Equation~\ref{eq:mag_eq}, which replaces second spatial derivative by a factor of $f^2$ and $\vscaled^2$. 
We replaced $R_F^2\simeq\lambda \aorb\sin i$, and we have additionally inserted a Wiener filter $W(f)$. Note that since this calculation is done in the Fourier domain, Equation~\ref{eq:model_mag} requires that variations in DM stay stationary within the relevant times.

For the fit, we take sections of the data in the egress $T\simeq\qty{65}{s}$, large enough to get a good estimate of the power spectra but small enough that the signal and noise are relatively stationary, and compare the power spectra of the model light curve $\mu_\text{model}$ and measured light curve $\mu$. More precisely, the power spectra of the model and measured magnifications are computed as
\begin{align}
    \Delta_{\mu}^2(f) = f\cdot\left|\fourier[\mu](f)\right|^2/T.
\end{align}

\subsubsection{Wiener filter}

\begin{figure*}
    \begin{center}
        \begin{adjustbox}{clip,trim=0.0cm 0cm 0.0cm 0cm, max width=.95\textwidth}
        \input{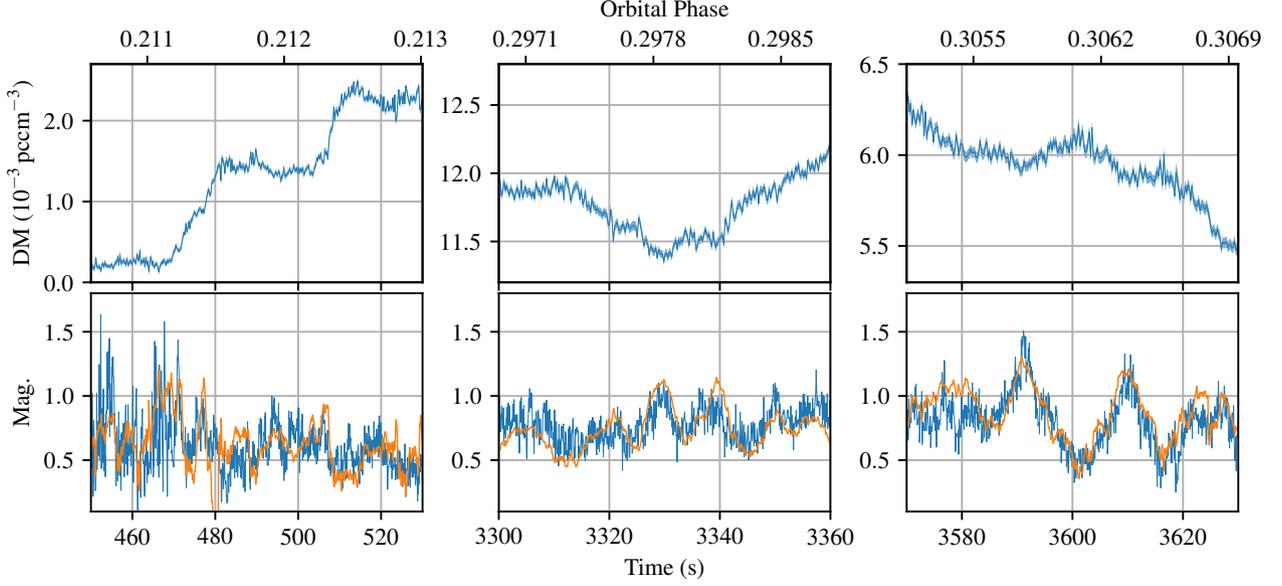}
        \end{adjustbox}
    \end{center}
    \vspace{-5mm}
    \caption{
    DM variations vs. observed and predicted magnifications at a few chosen locations in our data, showing the finer structure in DM variations. DM variations are averaged in time by a factor of 3 to 0.24-s for visual clarity. In the bottom panels, the blue line is the measured magnification averaged over 50 pulses, and orange is our model. Notice the peaks in model magnifications roughly correspond to local minima of DM variations, and vice-versa, as expected from Equation~\ref{eq:mag_eq}. The leftmost panels show a section of the ingress, and the middle and right panels show sections in the egress in the `optimal' fit zones. The right panel is a subsection of section C from Figure~\ref{fig:fit}.
    }
    \label{fig:optimal_dm_mag}
\end{figure*}

The Wiener filter $W(f)$ in Equation~\ref{eq:model_mag} serves two purposes. First, as previously discussed, DM measurements are significantly noisier than the magnification measurements, and taking derivatives of an already noisy time-series is an noise-amplifying process. The Wiener filter acts as an effective filter for the noise. Second, it is clear that there is some physical `effective smoothing scale' in the weak lensing regime, which can be seen from the turn-around in the power spectra in e.g. the third row of the left column in Figure~\ref{fig:magdmPS}. The Wiener filter mimics this smoothing process.
Physically, the smoothing scale is roughly the size of the scattering disk (see Section~\ref{sec:weak_manyimg} for discussions).

We use the Fourier domain form of the Wiener filter $W(f)=S(f)/(S(f)+N)$. The signal power $S$ is taken to be $S(f) = A f^{-3}$, a power-law that roughly describes large scale behavior of the DM power spectrum in the egress with the proportionality constant $A$ as a free parameter that will be fit for. The noise power $N$ is taken to be a constant white noise term for the section under consideration, obtained from the average noise floor at large $f$.

\subsubsection{Fitting procedure and results}
We logarithmically bin the power spectra above a cutoff frequency and fit the semi-logarithmically binned power spectra $\Delta_{\mu_\text{model}}^2(f)$ to $\Delta_{\mu}^2(f)$ with two parameters, $A$ and $\vscaled$,
using maximum likelihood. That is, we minimize the negative log-likelihood, taking the unbinned frequencies to be $\chi^2$-distributed with two degrees of freedom, and the errors $\sigma_i^2$ in the binned frequencies to be Gaussian-distributed with standard error of values in the bin,
\begin{align}
    -\log\mathcal{P}(A,\vscaled) = &\sum_{i=0}^{B-1} \log \left(\Delta_{\mu_{\text{model}},i}^2\right) +\frac{\Delta_{\mu,i}^2}{\Delta_{\mu_{\text{model}},i}^2} +\nonumber\\
    &\frac{1}{2}\sum_{i=B}^{N} \frac{\left(\Delta_{\mu,i}^2-\Delta_{\mu_{\text{model}},i}^2\right)^2}{\sigma_i^2} + \log\left(2\pi\sigma_i^2\right),
\end{align}
where $B$ is the cutoff threshold below which frequencies are not binned and $N$ the total number of frequency bins.
The errors of the two power spectra are combined in quadrature for the binned frequencies. 
Once appropriate $A$, and hence $W(f)$, and $\vscaled$ are fit for, the model magnifications can be computed using Equation~\ref{eq:model_mag} for the given section. The fitting procedure generally manages to recover the input velocity in simulations (see Appendix~\ref{sec:fittingsim} for details). 

There exists an `optimal' zone in the egress where this fit performs well both in frequency and time domain. 
That is, the model power spectrum $\Delta_{\mu_\text{model}}^2$ fits well to the measured power spectrum $\Delta_{\mu}^2(f)$; and the actual model time-series $\mu_\text{model}$, computed using Equation~\ref{eq:model_mag}, visually fits $\mu$ well (see Fig.~\ref{fig:fit} section C). 
In the optimal zone, the lensing timescale is $\simeq$ seconds, above the noise floor of the DM, but smaller than the timescales contaminated by absorption and ISS. In this zone, roughly between the orbital phase of 0.297 to 0.317, the magnification is well modeled linearly from DM. 
Figure~\ref{fig:optimal_dm_mag} shows a few small sections with DM and magnification variations where our fitting performs reasonably well. 
At later times, but before the strong regime, when the lensing timescale is $\lesssim\qty{1}{s}$, DM is noisy and although the power spectra still fit in the frequency domain, the correspondence is more statistical.
For example, in region 8, corresponding to the red curve on the third row, left panel of Figure~\ref{fig:magdmPS} peaking at $\sim\qty{1}{s}$, the DM is below its noise floor, i.e. DM measured on $\qty{1}{s}$ is noise-dominated. However, the magnifications inferred from DM still correlate with the measured magnification (see e.g. panel D in Fig~\ref{fig:fit} for the fit in the region and Section~\ref{sec:correlation} for a discussion on correlation). 

At earlier times, near the eclipse, since absorption, ISS, and the longer lensing timescales these all act on timescales of tens of seconds to minutes, fitting $\vscaled$ and $A$ is difficult due to their covariance. Instead of fitting for $\vscaled$ in those sections, we fix $\vscaled$, and only fit for the Wiener-filter coefficient $A$. This acts as a cross-check to the measured velocity for the egress.

Similarly, in the ingress, the timescales change too quickly to get a very good estimate on $\vscaled$, i.e., only $\simeq$ 1--2 min of the data is in the optimal zone. We use smaller sections of $\simeq \qty{40}{s}$. The `ingress spike', in particular, is not well fitted using our linear power spectrum fitting method.

We measure an ingress effective velocity of $\qty{954 \pm 99}{km.s^{-1}}$, and an egress effective velocity of $\qty{604 \pm 47}{km.s^{-1}}$ using 10 linearly-spaced bins and 13 log-linear-spaced bins ($B=10$, $N=23$). The ingress has a faster effective velocity than the egress, and the velocity throughout is significantly higher than expected from the eclipsing material comoving with the companion. We discuss the implications of this in Section~\ref{sec:interpretation_velocity}.

\subsection{Correlation between measured and predicted magnifications}
\label{sec:correlation}

To check the correlation between the model magnifications $\mu_\DM$ and the measured magnifications, we first compute the `intrinsic' magnification $\mu_\text{ecl}\equiv\mu/\mu_\text{bg}$ by dividing out the large scale variations $\mu_\text{bg}$ where it is sufficiently large (orbital phase $\gtrsim 0.286$). This also avoids spurious correlations that are not due to lensing by intra-binary materials. 
We compute the Pearson-$r$ cross-correlation in a 180-s rolling window (2240 samples per window) using
\begin{align}
    r(t)=\frac{\left\langle(\mu_\DM-\langle\mu_\DM\rangle)(\mu_\text{ecl}-\langle\mu_\text{ecl}\rangle)\right\rangle}{\sqrt{\var(\mu_\DM)-\err(\mu_\DM)^2}\sqrt{\var(\mu_\text{ecl})-\err(\mu_\text{ecl})^2}}
\end{align}
where in the denominator, we correct the data variances by the expected error due to measurement error on $\DM$ or intrinsic noise for $\mu_\DM$ and $\mu_\text{ecl}$, respectively.

To estimate $\err(\mu_\DM)^2$, we employ Monte-Carlo methods. More precisely, we simulate 1000 realizations of $\mu_{\DM_{\text{sim}}}$ using $\DM_\text{sim}(t) = \DM(t)+G_1(t)\,\sigma_\DM(t)$, $v_\text{scaled, sim}(t) = \vscaled(t)+G_2(t)\,\sigma_{\vscaled}(t)$, and $A_\text{sim}(t)=A(t)+G_3(t)\,\sigma_A(t)$
where $G_i(t)$ are drawn from a standard Gaussian distribution, i.e. each is a single realization of white noise. 
We use the standard deviation across the 1000 sims at a certain time $t$ as the model error of $\mu_\DM$ at that time. The $\err(\mu_\DM)^2$ in a 180-s window is then the mean squared error of $\mu_\DM$ within the 180-s window. 
To estimate $\err(\mu_\text{ecl})^2$, we compute the power spectral density of $\mu_\text{ecl}$, and set $\err(\mu_\text{ecl})^2$ as the average power of high frequency components, where the noise floor, arising from the intrinsic pulse-to-pulse variations, is apparent. Alternatively, this term can be inferred from the out-of-eclipse RMS variations.
Once we have the error terms, the error on $r(t)$ is estimated using with 10000 pairs of bootstrapped samples in each 180-s window.

\begin{figure}
    \begin{center}
        \begin{adjustbox}{clip,trim=0.4cm 0cm 0.0cm 0cm, max width=0.45\textwidth}
        \input{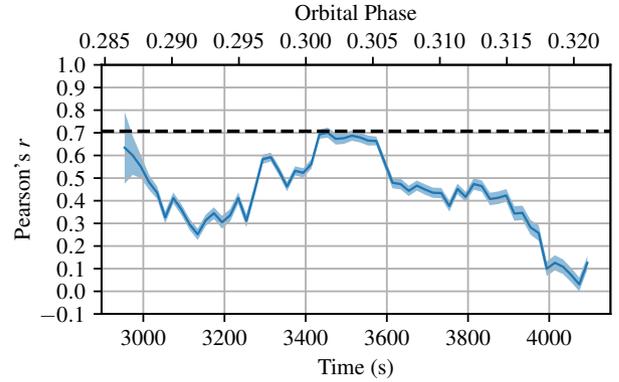}
        \end{adjustbox}
    \end{center}
    \vspace{-5mm}
    \caption{
    Pearson correlation coefficient $r(t)$ between $\mu_\DM$, the model, and $\mu_\text{ecl}$, the measured magnifications due to the eclipse in the egress of the eclipse. Each point on $r(t)$ is computed from a 180-s window centered on the point, and the shaded color indicate the bootstrapped error bar. The black dashed line is drawn at $1/\sqrt{2}$, the expected correlation between a 1D model and a 2D isotropic lens. 
    }
    \label{fig:pearsonr}
\end{figure}

Using Equation~\ref{eq:mag_eq}, and an analogous equation for lensing in 2D \citep{linDiscoveryModellingBroadscale2021}, we can evaluate the expected correlation coefficient between a 1D model and a 2D lens (the expected correlation for a 1D model and a 1D lens, is, of course, $r=1$ in the absence of noise). Assuming an isotropic Gaussian random field for DM, one can compute that the expected $r\lesssim 1/\sqrt{2}$ (see Appendix~\ref{sec:correlation_appendix} for details). We show this upper bound for correlation, as well as $r(t)$, in Figure~\ref{fig:pearsonr}. 
The correlation is significantly positive throughout, except for possibly a few 180-s windows near orbital phase of $0.320$. 
In contrast, two uncorrelated timeseries are expected to have a correlation of $0$, with standard error of $\sim 1/\sqrt{2240}$, where 2240 is our number of samples per window.
Moreover, the correlation is consistent with the $1/\sqrt{2}$ bound in the best regions. 

Noticeable difference from the expected correlation can be seen at orbital phases between 0.288 to 0.297 and 0.315 to 0.321. 
In addition to errors being underestimated, 
at earlier times before the orbital phase of 0.297, there is measurable scattering, hence necessarily multiple imaging, causing the measured DM to be `smoothed out' over $\gtrsim \qty{15}{s}$ scales compared to the physical line of sight column density as discussed in Section~\ref{sec:weak_scattering}, which reduces correlation. 
At later times for orbital phase 0.315 onwards, the presence of strong lensing, implying that Equation~\ref{eq:mag_eq} is no longer valid, can cause the correlation to drop.

\subsection{Ingress brightness spike}
\label{sec:ingress_spike}
The ingress brightness spike is a feature seen in the magnification curve in ingress at the orbital phase $\sim 0.215$ in Figure~\ref{fig:measurements_58261} and Figure~\ref{fig:fit}, where absorption is apparent. 
The 2-s scatter-corrected magnification goes from $\sim 0.5$ before the spike, to $\sim 1$ at the peak without exceeding 1, and drastically drops to $\lesssim 0.2$ after the peak, with a full width at half max of $\sim\qty{14}{s}$. 
Interestingly, the scattering timescale increases significantly right after the ingress spike, from $\lesssim\qty{20}{\us}$ to $\sim\qty{60}{\us}$.
It appears to be roughly achromatic, similar to the rest of the `weak lensing' regime. 

Initially, the ingress spike appears to be easy to model: it preceeds a rapid increase in DM, from $\qty{2e-3}{pc.cm^{-3}}$ to $\qty{5e-3}{pc.cm^{-3}}$ in roughly $\sim\qty{30}{s}$. Such an increase in DM naturally leads to a caustic or near-caustic crossing \citep[see, e.g.][]{cleggGaussianPlasmaLens1998}. 
Although modeling from observed DM succeeds at other places in the weak lensing regime, it fails here: the sharp rise in DM does not produce a brightness spike at the right location on the screen. 

The refractive shift due to the sharp DM increase is 
\begin{align}
    \Delta x\simeq \qty{6070}{km} \frac{\Delta\DM}{\qty{3e-2}{pc.cm^{-3}}}\frac{\qty{30}{s}}{\Delta t}\frac{\qty{360}{km.s^{-1}}}{\veff},
\end{align}
which, modulo order 1 factors, corresponds to $\gtrsim\qty{10}{s}$ shift, that is, the real position of the source and the observed image may be offset by greater than \qty{10}{s} in crossing time. At the same time, the scattering disk size corresponding to a scattering time of $\sim\qtyrange{20}{60}{\us}$ is $\sim\qtyrange{4500}{8000}{km}$, also corresponding to a $\sim\qtyrange{10}{20}{s}$ crossing time. These effects make modelling the spike directly from observed DM more complicated, since the observed DM is already influenced by lensing. 

\begin{figure}
    \begin{center}
        \begin{adjustbox}{clip,trim=0.4cm 0cm 0.0cm 0cm, max width=0.45\textwidth}
        \input{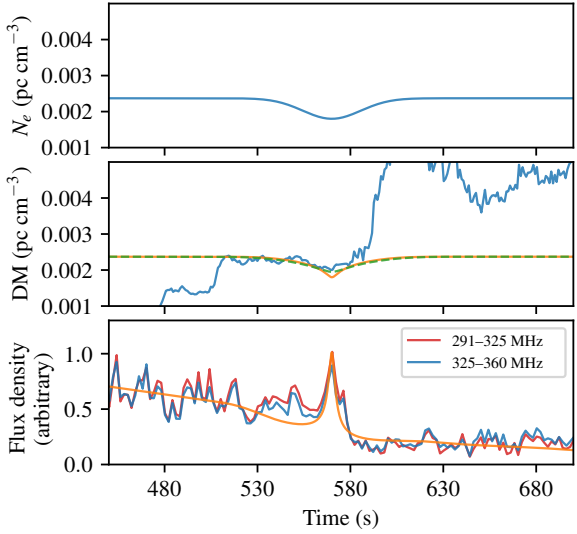}
        \end{adjustbox}
    \end{center}
    \vspace{-5mm}
    \caption{
    Modelling the `ingress spike'. Top panel shows the lens plane column density $N_e$. Middle panel shows the source plane DMs: blue is the observed DM rebinned to 1 second, orange is the `lensed' DM, and green dashed is the lensed DM smoothed by a 10-s gaussian to mimic the effect of the scattering. The green, dashed curve roughly reproduces the dip seen in the observed DM.
    The bottom panel shows the scatter-corrected flux density at the bottom and top halves of the band as the red and blue lines, and the model magnification as the orange line. The model magnification roughly reproduces the ingress spike in amplitude, location, and duration, but not its achromaticity, as discussed in Section~\ref{sec:ingress_spike}.
    }
    \label{fig:ingress_spike}
\end{figure}

Instead, we put in a simple analytical 1D column density model and solve the lens equation to see whether we can reasonably reproduce the brightness spike, without fitting.
We choose
\begin{align}
    N_e(x)=N_{e,0} + N_{e,1}\exp\left[-\left(\frac{x-x_1}{\sigma_1}\right)^2\right],
\end{align}
where the exponential term models the minor dip in DM before the steep increase, and solve the lens equation \ref{eq:lens_eq} assuming $\vscaled=\qty{288}{km.s^{-1}}$. The top panel of Figure~\ref{fig:ingress_spike} shows the lens plane column density, $N_e(x)$.

The source plane, or observed $\DM(x_s)$ and magnification $\mu(x_s)$ are shown as the orange line in the second and bottom panel of Figure~\ref{fig:ingress_spike}. The spatial coordinates have been converted back into time coordinates using orbital velocity of \qty{360}{km.s^{-1}}. In the second panel, we also show $\DM(x_s)$ in green smoothed with a 10-s gaussian to mimic the effect of the scattering. 
The smoothed DM curve roughly reproduces the observed DM at the right location, including the dip at $t\simeq\qty{570}{s}$, and the magnification curve roughly reproduces the ingress spike. The assumed velocity of \qty{288}{km.s^{-1}} is significantly smaller than the velocities measured in the ingress with weak lensing, but it is plausible for a single, large anisotropic lens with an orientation $\alpha\simeq \ang{37}$ to cross the line of sight.

In practice, it is difficult to find a column density profile that satisfies the achromaticity, location, duration, and amplitude of the observed spike. For example, if the spike is solely due to geometric lensing (i.e., not the lack of absorption), it has a magnification of $\sim 2$. Expanding the first line of Equation~\ref{eq:mag_eq}, 
\begin{align}
    \label{eq:1dchromaticity}
    \mu(\lambda_0+\delta\lambda)\simeq \mu + 2\mu(\mu-1)\frac{\delta\lambda}{\lambda_0}.
\end{align}
For $\mu=2$ and a fractional bandwith of $\delta\lambda/\lambda_0\simeq0.15$, the second term is $\sim 0.6$, and therefore any 1D model will be chromatic over our band. A transverse dimension can reduce the chromaticity due to the additional shear term, modifying Equation~\ref{eq:1dchromaticity}
\begin{align}
    \label{eq:2dchromaticity}
    \mu(\lambda_0+\delta\lambda)\simeq \mu + 2\mu\left(\mu\left(1-\kappa^2+\gamma^2\right)-1\right)\frac{\delta\lambda}{\lambda_0},
\end{align}
where $\kappa$ and $\gamma$ are the convergence and shear, respectively \citep[see e.g.][]{tuntsovDynamicSpectralMapping2016}. 

Although the large, sudden, increase in column density is unlikely the main cause of the brightness spike due to its location, it may contribute to the some of the increase in flux density seen, especially near the right edge of the spike. Additionally, it likely contributes to the sudden decrease in brightness from $\sim 0.5$ to $\sim 0.2$ after the spike, since it would still act as a diverging lens, deflecting light out of the line of sight.

\section{Interpretations of scaled velocity}
\label{sec:velocity}

\label{sec:interpretation_velocity}
The observed velocities, both in ingress and egress, are $\sim$~2--3 times higher than naively expected from the lensing material co-moving with the companion at the relative orbital velocity of $\gtrsim \qty{360}{km.s^{-1}}$\citep{vankerkwijkEvidenceMassiveNeutron2011}. This cannot be explained solely by a lower inclination $i\lesssim\ang{30}$ of the system, since it implies a pulsar mass of $\gtrsim \qty{10}{\msun}$. Similarly, a mass ratio of $\gtrsim 130$ is also not possible. 
Since we directly measure $\vscaled$, we are left with a few possibilities: a lensing angle $\alpha$ offset from the relative motion, the lensing material being significantly closer to the pulsar, or a significant flow velocity.

\begin{figure}
  \begin{center}
    \begin{adjustbox}{clip,trim=0.0cm 0cm 0.0cm 0cm, max width=0.45\textwidth}
    \input{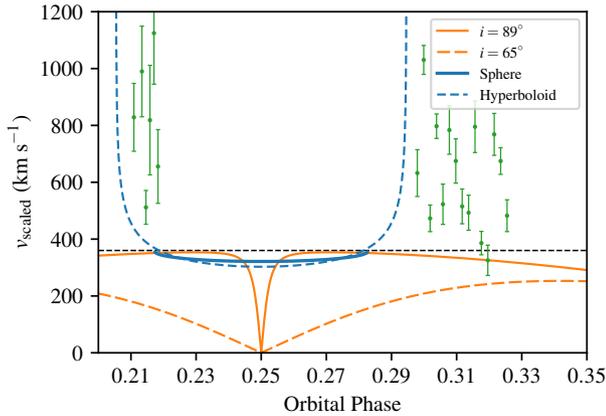}
    \end{adjustbox}
    \vspace{-2em}
  \end{center}
\caption{
The blue lines show $\vscaled$ assuming edge-on inclinations and spherical (dashed) and hyperboloidal (solid) geometry of the eclipse, and the orange lines show $\vscaled$ assuming a changing $\alpha$ at two different inclinations of the system (See Section~\ref{sec:interpretation_velocity}). 
The green points show the measured velocities from Figure~\ref{fig:fit}.
The dashed line is the projected orbital velocity $\vorb\sin\,i\simeq\qty{360}{km.s^{-1}}$.
For the sphere, we take the size of the sphere to be the size the eclipse. For the hyperboloid, we assume a rotationally symmetric hyperboloid, and take the circle at an orbital separation of $\sim \qty{2.7}{\rsun}$ to be the size of the eclipse, similar to the one shown in Figure~\ref{fig:system_geometry}. The scaled velocity is then scaled by the distance from the pulsar to intersection of the line of sight and the sphere/hyperboloid. 
The steep increase at orbital phase of $0.21$ and $0.29$ is due to the requirement that the sphere/hyperboloid be co-rotating.
At two different inclinations ($i=\ang{65}$ in dashed orange and \ang{89} in solid orange) of the system, we show the effect of $\alpha$ changing as the orbit progresses on $\vscaled$, assuming that the line of scattered images is radial from the companion as described in Section~\ref{sec:interpretation_velocity}.
These geometries, which only depend on the orbital motions, cannot explain the large velocities observed without an additional outflow velocity. 
}
\label{fig:velocity_models}
\end{figure}

For the first case,
a misaligned lensing axis with the direction of motion, with an angle $\alpha$ between the two does not explain the larger velocity, since the misalignment angle can only decrease velocity, as per Equation~\ref{eq:effective_velocity}. 
We briefly describe the model for completeness. 
If the eclipsing material is spherically distributed around the companion, the gradient of the projected column density, and hence the lensing direction, would be radial from the companion. Hence, an inclination $i$ of the system that is different from \ang{90} will cause a misalignment of the motion vector and the lensing angle. 
This can, in principle, be used to independently constrain the inclination angle $i$. 
The angle $\alpha$ reaches the maximum of $\ang{90}$, and turns around at the orbital phase of $0.25$, and hence the effective velocity momentarily reaches 0 there. 
We show the effect of $\alpha$ varying through the orbit at different inclinations of the system in Figure~\ref{fig:velocity_models}.

To estimate the effect of changing the source-lens distance $\dsl$, we consider two simple geometric models: spherically symmetric material around the companion, and a hyperboloidal shock around the companion, so that the distance from the pulsar to the surface of the sphere or hyperboloid changes over the eclipse. 
We assume an edge-on inclination, take the eclipse to be $\sim \qty{1.1}{\rsun}$, and that the material is co-rotating with the companion. 
For the sphere, we take the size of the sphere to be the size the eclipse. For the hyperboloid, we assume a hyperboloid of rotation, and take the circle at an orbital separation of $\sim \qty{2.7}{\rsun}$ to be the size of the eclipse.
The resulting $\vscaled$ is shown in Figure~\ref{fig:velocity_models}.

Finally, a large projected flow velocity could contribute to the observed velocity. 
In the ingress, the flow velocity is moving in the same direction as the companion, and thus the sign in front of $\Vflow$ Equation~\ref{eq:effective_velocity} would be modified to be positive. 
If the lensing is happening roughly at the projected binary separation, and $\alpha\simeq 0$, we can solve for the flow velocity assuming that the effective velocity to be our measured ingress scaled velocity, $v_{\text{flow},\perp}\sim \qty{950}{km.s^{-1}}-\qty{360}{km.s^{-1}} \sim \qty{590}{km.s^{-1}}$.
Conversely, in the egress, the flow velocity is opposite of the motion of the companion, and the flow velocity in that case is $v_{\text{flow},\perp}\sim \qty{600}{km.s^{-1}} + \qty{360}{km.s^{-1}} \sim \qty{960}{km.s^{-1}}$. 
The larger projected velocity in the egress is plausible due to the asymmetry of the eclipse. Additionally, the large flow velocity relative to the orbital velocity means that the lensing material's motion is retrograde in the egress. 
From geometry, the $\Vflow$ near the eclipse is likely close to radial (see Figure~\ref{fig:system_geometry}), and thus these projected values are compatible with the companion wind speed directly measured from X-ray observations, $10.8_{-3.5}^{+6.0}\cdot \vorb$ assuming an inclination of \ang{90} of the system \citep{kandelXMMNewtonObservesIntrabinary2021}.
Finally, since the flow velocity is of the same order as the orbital velocity, the lenses should reconfigure on the order of the orbital period, and thus we are not seeing the same lenses from orbit to orbit.

This is an independent way to bound the outflow velocity from the companion, since we constrain the outflowing velocity perpendicular to the line of sight, and hence give a lower bound of the companion mass-loss rate. The mass-loss rate is $\dot{m}_c\simeq\pi R_E^2 m_\text{pr} n_\text{e} \vflow$, where $R_E$ is the half-width of the eclipse, $m_\text{pr}$ the proton mass, and $n_\text{e}$ the electron density \citep{thompson+94}. Our eclipse width is $\sim \qty{1.11}{\rsun}$. At an inclination of \ang{65}, this corresponds to a chord across a circle of diameter $\qty{2.75}{\rsun}$ \citep[cf.][]{polzin+20}. Estimating $n_\text{e} \simeq \qty{0.025}{pc.cm^{-3}}/\qty{2.75}{\rsun} = \qty{4e5}{cm^{-3}}$, and using our egress outflow velocity of $\qty{960}{km.s^{-1}}\leq \vflow$, the companion mass-loss rate is $\geq\qty{3e-14}{\msun.yr^{-1}}$, implying a companion evaporation timescale of $\leq 80$ Hubble times for a companion mass $m_c\simeq \qty{0.035}{\msun}$.

This mass loss rate is a very conservative bound, since in addition to $v_{\text{flow},\perp}\leq \vflow$, the particle density is estimated from the measured excess DM near the eclipse. If the inclination of the system is indeed $\simeq\ang{65}$, we are only seeing the outer edge (i.e. lower density contours) of the `sphere' of eclipse. Moreover, the average electron density inside the eclipsing sphere may be much higher than the above estimate of $\sim\qty{4e5}{cm^{-3}}$ if the column density continues like a power-law into the eclipse. These two effects can combine to contribute 1--2 orders of magnitude to the electron density in the eclipse, and hence the companion mass-loss rate, implying that the companion may evaporate within a Hubble time.

\section{Summary and Interpretation of Lensing}
\label{sec:interpretation}
To summarize the observed diversity in lensing behavior, we provide here the list of key properties:

\begin{enumerate}
    \item When line of sight passes closest to eclipse, a `slow' and weak lensing regime, with timescales ranging from minute-long close to the eclipse to $\sim$ few tenths of a second near the end of this phase. The modulations are achromatic, and the amplitudes stays relatively constant (variance $\sim 0.4$). This region is well modeled directly from DM. Additionally, during times closest to the eclipse, there is significant absorption, as well as measurable scattering, from $\sim\qty{100}{\us}$ to $\sim\qty{10}{\us}$, which is roughly the smallest measurable scattering time using our profile matching method.
    \item As the line of sight moves away from the eclipse, strong lensing suddenly `turns on', within $\lesssim\qty{10}{s}$ in the ingress and $\lesssim\qty{25}{s}$ in the egress. Once strong lensing is on, it stays on until the pulsar is no longer lensed by the intrabinary material.
    \item In the strong lensing region, very chromatic, high modulation amplitude lensing events, called `extreme lensing' in \citet{mainPulsarEmissionAmplified2018}, last $\sim$ 5-10 pulses in the egress, and somewhat shorter in the ingress. They also sometimes appear to `sweep' in time and frequency.
    \item Farther from the eclipse, strong lensing events get less and less frequent, but the weak lensing regime does not reappear.
    In the ingress, very infrequent events still exist somewhat far from the eclipse. 
    Using method outlined in Appendix B of \citet{li+19} to distinguish lensed pulses from giant pulses, we find the first lensing event with $\mu>4$ appears at the orbital phase of $\sim 0.162$, with roughly 6 events in the subsequent \qty{3}{min}, or once every $\sim 18600$ pulse rotations.
    In the egress, this part is unfortunately cut off by the observation in the day analyzed in this paper.
    \item Far away from the eclipse, the pulsar returns to its usual brightness, as shown by \citet{mainPulsarEmissionAmplified2018}.
\end{enumerate}

These happen asymmetrically in duration in the ingress and egress, with each stage lasting much longer during the latter. The transition from the measurably scattered region to the non-measurably scattered region seems to be uncorrelated with the total DM, with the transition happening at a much lower DM in ingress $\sim$~(\qtyrange{1e-4}{1e-3}{pc.cm^{-3}}), than in egress $\sim$~(\qtyrange{7e-3}{1e-2}{pc.cm^{-3}}). 
On the other hand, the transition from weak to strong lensing happens in both the ingress and egress when DM decreases to $\lesssim\qty{3e-4}{pc.cm^{-3}}$.

\subsection{Difference in ingress and egress transitions}

The difference between DM in transition to measurable scattering may be due to a difference in the screen distance between ingress and egress. 
Scattering time is
$\tau=\frac{\dl\ds}{\dsl} \frac{\theta^2}{2c}$,
and scattering disk size $\theta$ is $\theta = \frac{\sqrt{\langle\mathbf{x}_i^2\rangle}}{\dl}$,
where $\mathbf{x}_i^2$ are the location of the images, i.e. solution to the lens equation~\ref{eq:lens_eq} \citep{gwinnInterstellarOptics1998}. 
Thus, 
\begin{align}
\tau\propto \frac{\left\langle(\mathbf{x}_i)^2\right\rangle}{\dsl}.
\end{align}
That is, for the same scattering time, a screen farther from the source has a larger scattering disk size, the lenses have to deflect from farther away, and hence more material is typically needed.

\subsection{Refractive interpretation}
\label{sec:geomopt_interpretation}
We propose a self-consistent, plausible scenario to the lensing observed in \psr. 
It is clear that multiple imaging occurs near the eclipse, since we see clear scattering.
Scattering can be due to small scale ($\ll R_F$) structures, however, we propose instead that they are due to larger scale lenses which refract light from farther away. 
The collection of the many refractive images can emulate scattering.
As we progress from the near eclipse to farther away, the measured scattering time decreases. 
Since DM is simultaneously decreasing, the decrease in scattering time can be attributed to, for example, decreasing lensing strength of the individual lenses and density of lenses, which leads to images refracted from less far away. 
The weak lensing regime is then due to the average effect of these many images, and is thus not a `real' weak regime, but rather an `effective' one (see also discussions in Section~\ref{sec:weak_manyimg}). 
This also explains the strong lensing regime at lower DM, as well as the transition to quiescence: the lensing strength/density of lenses have decreased to a point where only a few images contribute coherently, leading to chromatic events with high peak magnifications. 
Finally, this suggests that the strong lensing events are not purely due to wave effects. The fine time-frequency structure due to pure wave effects near caustics, like in simulations by \citet{li+19} and \citet{melroseScintillationRadioSources2006}, may be smeared over by the finite frequency resolution of the telescope and the finite time resolution.

To examine this quantitatively, we consider the lensing parameters discussed in \citet{jowRegimesAstrophysicalLensing2022}, who suggest that refractive/geometric optics may be applicable to most of the parameter space in lensing. 
More specifically, for a single lens, the relevant parameter that can be used to estimate the applicability of geometric optics is not the Fresnel scale $R_F$, but rather the dimensionless parameter 
\begin{align}
\label{eq:epsilon}
    \epsilon&=\lambda r_e N_e \nonumber\\
    &\simeq\num{8.7e4}\frac{\lambda}{\qty{1}{m}}\frac{N_e}{10^{-3}\,\unit{pc.cm^{-3}}},
\end{align}
which depends only on the observational wavelength $\lambda$ and the maximum column density of the lens $N_e$. When $\epsilon\gg 1$, geometric optics can be applied.
In addition, the convergence of the lens
\begin{align}
\label{eq:kappa}
    \kappa&\simeq\frac{\dsl\dl}{\ds}\frac{1}{r_\ell^2}\frac{\lambda^2 r_e}{2\pi}N_e\nonumber\\
    &\simeq\num{9.54}\frac{\dsl}{\qty{1}{kpc}}\frac{\dl}{\qty{1}{kpc}}\left(\frac{\ds}{\qty{2}{kpc}}\right)^{-1}\left(\frac{r_\ell}{\qty{1}{AU}}\right)^{-2}\left(\frac{\lambda}{\qty{1}{m}}\right)^{2}\frac{N_e}{10^{-3}\,\unit{pc.cm^{-3}}},
\end{align}
where $\dl$ is the distance to the lens, $\ds$ distance to the source, and $r_\ell$ the lens size, determines whether the lensing is strong or weak, since caustics appear when $\kappa \gtrsim 1$ \citep[cf.][]{cordesLensingFastRadio2017}. In the second line of the above equations, we inserted values for a typical ISM lens.

To estimate these parameters in \psr, we need to estimate the size and excess DM of lenses. 
We note that \citet{jowRegimesAstrophysicalLensing2022}'s conclusion comes from an analysis of a single lens, and here we may be observing the combined effects of many lenses, and hence would be measuring an `effective' $\epsilon$ and $\kappa$ for the many lenses. 
Since $\epsilon$ simply depends on $N_e$, multiple lenses may contribute additively.
The dependence is more complicated for the convergence $\kappa$.
Indeed, lensing seen in \psr may be useful to guide theoretical investigations. 

In the weak, linear, lensing regime, we take the autocorrelation timescale of magnifications (see panel f) of Figure~\ref{fig:measurements_58261}) times the relative orbital velocity \qty{360}{km.s^{-1}} as the lens size. 
On the other hand, in the strong regime, larger lenses are expected to lens for a shorter time due to focusing. 

Using the perfect lens approximation, the full-width at half max timescale of strongly lensed events are \citep[cf. Equation~18 in the appendix of][]{mainPulsarEmissionAmplified2018}
\begin{align}
    \Delta t_\text{FWHM}\simeq \frac{0.45 R_F^2}{r_\ell\,v},
\end{align}
taking $\Delta t_\text{FWHM}$ to be the measured autocorrelation timescale, $v$ as the mean egress scaled velocity, and $R_F\simeq \qty{40}{km}$ to be at the projected binary separation.
The parameter $r_\ell$ in the perfect lens approximation is the size of the coherently added area, and not a physical size of the lens, but well approximates the size of the lens for bright events.

For $N_e$, the DM power spectrum of in the egress is well estimated (see third row, right column of  Figure~\ref{fig:magdmPS}), and fits a power-law index of $\sim -3$. Assuming that $N_e\simeq \DM$, and that the functional form of the power spectrum of DM fluctuations is indeed a power-law with index $-3$, and converting from time to space using the orbital velocity, we can look up the rms fluctuations in DM at the scale of the estimated lens sizes $r_\ell$. We take the rms fluctuations to be $N_e$. 

\begin{figure}
    \begin{center}
        \begin{adjustbox}{clip,trim=0.4cm 0cm 0.0cm 0cm, max width=0.45\textwidth}
        \input{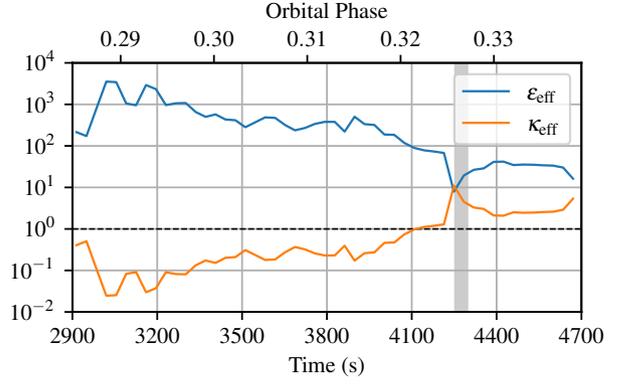}
        \end{adjustbox}
    \end{center}
    \vspace{-5mm}
    \caption{
    Estimated lensing parameters $\epsilon_\text{eff}$ and $\kappa_\text{eff}$ versus time in the egress. $\epsilon_\text{eff}$ stays well above unity for the entirely of the egress, indicating that lensing can be well modeled by geometric optics. $\kappa_\text{eff}$ is below unity at earlier times, and jumps to above significantly above unity for $t\gtrsim \qty{4200}{s}$, roughly consistent with the transition between observed weak and strong lensing regime, shown as the grey shaded region.
    }
    \label{fig:epsilon_kappa}
\end{figure}

From there, we can get both $\epsilon_\text{eff}$ and $\kappa_\text{eff}$ in the egress. Figure~\ref{fig:epsilon_kappa} show the two lensing parameters versus time in the egress. Indeed, $\epsilon_\text{eff}$ stays well above unity, implying that geometric optics apply, and $\kappa_\text{eff}$ goes from below unity to above unity at $t\gtrsim\qty{4200}{s}$, exactly at the observed transition of weak to strong lensing regimes. This matches the single-lens prediction by \citet{jowRegimesAstrophysicalLensing2022}. Therefore, from typically measured observables, such as flux, DM, and scattering, we can predict the transition region to strong lensing.

This does not imply that strong lensing always occurs in the region between the effective weak lensing region and quiescence in an eclipsing system. The effective $\kappa$ may stay well below 1, for example, if lens sizes are large. In \psr, $N_e\simeq\qty{4e-7}{pc.cm^{-3}}$ and $r_\ell\simeq\qty{60}{km}$ in the strong lensing region. 
If, instead, lenses were $\sim\qty{200}{km}$, $\kappa$ could stay well below unity for the whole of the non-quiescent region, and observable strong lensing may not `turn on'. 

\subsection{Apparent weak lensing, but many images?}
\label{sec:weak_manyimg}
Similar to PSR~J2051$-$0827 \citep{linDiscoveryModellingBroadscale2021}, the light curve in \psr, well-modeled by weak lensing using Equation~\ref{eq:mag_eq}, occurs simultaneously with observable scattering. 
Weak lensing is often understood to imply a single, coherent weakly-magnified image, yet observed scattering implies the presence of many images.
Clearly, these cannot be simultaneously true. 
Which of the assumptions breaks down? 
The most fragile of these is inferring that the weak lensing implies a single, coherent image. 
Indeed, if the broad-scale DM structure can focus or defocus many individual rays in a way that mimics `weak lensing' (and hence, produce $\epsilon_\text{eff}$ and $\kappa_\text{eff}$ discussed above), then all observations are consistent. 
This work, as well as work by Jow et al. (in prep.) extend \citet{jowRegimesAstrophysicalLensing2022} suggesting that this is indeed the case. 
Indeed, simulations suggest that large-scale structure can mimic weak lensing even in the presence of many images created due to small-scale structures. 

This would immediately imply that the caustics, which exists for a smooth plasma lens at low enough frequency, predicted by \citet{linDiscoveryModellingBroadscale2021} in PSR~J2051$-$0827, never actually occurs for lower frequencies, contrary to the naive expectation for single-imaged weak lensing. 
Such caustics will always be smeared out by the averaging over the many images, since the `scattering strength' (or, the number of images) also increases towards lower frequencies. 

Conversely, we predict strong lensing at higher frequencies at the apparently weakly lensed regions. 
Consequently, the transition region must occur closer to the eclipse at higher frequencies. 
This prediction can be tested by observing pulsars exhibiting either apparent weak and strong lensing using many of the newer or upcoming ultra-wideband (UWB) receivers. 
Additionally, although the `weak lensing'-modeled light curve and the real light curve from multiple imaging are close in preliminary simulations, the inferred velocities could be influenced by differences between the two. We leave such investigation for the upcoming paper by Jow et al. (in prep.).

\section{Conclusions and Further Work}
\label{sec:summary}
In this paper, we investigated lensing around the eclipse of \psr, and showed that light curve modulations due to weak lensing can be predicted from the measured DM fluctuations using a two-parameter linear 1D lensing model. 
This confirms and improves on \citet{linDiscoveryModellingBroadscale2021} by providing a measure of the effective velocity of the lensing material as a function of pulse phase. 
The variable velocity is consistent with a significant flow velocity of the lensing material. 
Additionally, we showed variable DM, scattering, and absorption by intrabinary material. 

Lensing in \psr transitions relatively suddenly from an apparently weakly lensed, scattered regime to a strong lensing regime when the total dispersion measure drops below a critical value. 
This is likely due to the eclipsing material consisting of large `clumps' of plasma, which can individually cause strong lensing. 
When there are many clumps close to the line of sight (i.e. higher DM regions), light can be deflected from farther away into the line of sight, which may be interpreted as a scattering tail of the pulse. 
Contrary to expectations from single-imaged weak lensing, we expect strong lensing at higher frequencies, and no observable caustics at lower frequencies. This can be tested by observing pulsars exhibiting lensing with UWB receivers.
Moreover, the geometric optics regime may be enough to understand lensing in \psr.

Our analysis can be applied to other eclipsing pulsars in binaries with DM variations or observed lensing, such as the recently discovered PSR~J1720$-$0533 \citep{wangUnusualEmissionVariations2021}, or B1744$-$24A, which also exhibits strong lensing \citep{bilous+19}. 
In addition, the `laboratory-like' conditions in \psr may be useful to study pulsar lensing in the ISM, as well as lensing of FRBs. 
In either case, a positive correlation between the second derivative of DM and flux density can demonstrate lensing. 
Since the second derivative may be difficult to estimate in the ISM, DM itself may be correlated with flux density, but more care needs to be taken to avoid spurious (anti-)correlation due to long term trends. 
We leave such studies for the future.

\section*{Acknowledgements}
We thank the anonymous referee whose comments helped to significantly improve the presentation of this paper. 
We thank the Scintillometry group at the University of Toronto for general discussions. 
FXL thanks Luke Pratley for helpful discussions on the correlation coefficient. 
ULP receives support from Ontario Research Fund—research Excellence Program (ORF-RE), Natural Sciences and Engineering Research Council of Canada (NSERC) [funding reference number RGPIN-2019-067, CRD 523638-18, 555585-20], Canadian Institute for Advanced Research (CIFAR), the National Science Foundation of China (Grants No. 11929301), Thoth Technology Inc, Alexander von Humboldt Foundation, and the National Science and Technology Council (NSTC) of Taiwan (111-2123-M-001 -008 -, and 111-2811-M-001 -040 -).
The analysis made extensive use of the \texttt{scipy}, \texttt{numpy} \citep{2020SciPy-NMeth}, and \texttt{astropy} \citep{astropy:2013,astropy:2018} packages.

\section*{Data availability}
The data underlying this article is available upon reasonable request to the corresponding author.

\bibliographystyle{mnras}
\bibliography{bib}

\begin{appendix}

\section{Correlation}
\label{sec:correlation_appendix}
We derive the correlation for a 1D model when the physical lensing is due to a 2D system. Suppose the column density on the lens plane is $N_e(x,y)$. The lens potential is 
\begin{align}
    \psi(x,y)=\frac{\dsl}{\ds\dl}\frac{\lambda^2 r_e}{2\pi} N_e(x,y),
\end{align}
and the magnification is
\begin{align}
    \mu_\text{2D}&=\left(1-\left(\psi_{xx}+\psi_{yy}\right)+\psi_{xx}\psi_{yy}-\psi_{xy}^2\right)^{-1}\nonumber\\
    &\simeq 1+\left(\psi_{xx}+\psi_{yy}\right)-\left(\psi_{xx}\psi_{yy}-\psi_{xy}^2\right)
    ,
\end{align}
where subscripts indicate derivatives, and in the second line we expand for weak lensing. We only observe a 1D cross section of $N_e$. If the deflection due to lensing is sufficiently small, and smoothing due to scattering and multi-path lensing is not too significant, the measured column density $\DM(x)\simeq N_e(x,y=0)$, where we take $y=0$ to be the sampling axis. Hence we construct a 1D model magnification
\begin{align}
    \mu_\text{1D}\simeq 1+\psi_{xx}.
\end{align}

Now consider the correlation between $\bar\mu_\text{2D}=\mu_\text{2D}-1$ and $\bar\mu_\text{1D}=\mu_\text{1D}-1$, where we subtract the mean value of unity,
\begin{align}
    r_{\bar\mu_\text{1D},\bar\mu_\text{2D}}=\frac{\left\langle\bar\mu_\text{1D}\bar\mu_\text{2D}\right\rangle}{\sqrt{\left\langle(\bar\mu_\text{1D})^2\right\rangle}\sqrt{\left\langle(\bar\mu_\text{2D})^2\right\rangle}}.
\end{align}
Assuming a gaussian field $\DM(x,y)$ with zero mean, $\psi_{xx}$, $\psi_{yy}$, and $\psi_{xy}$ are independent, and have zero mean. Therefore,
\begin{align}
    \left\langle\bar\mu_\text{1D}\bar\mu_\text{2D}\right\rangle=\left\langle(\bar\mu_\text{1D})^2\right\rangle=\left\langle(\psi_{xx})^2\right\rangle\equiv\gamma^2,
\end{align}
and the 2D auto-correlation term is
\begin{align}
\label{eq:autocorr2d}
    \left\langle(\bar\mu_\text{2D})^2\right\rangle=\left\langle(\psi_{xx})^2\right\rangle+\left\langle(\psi_{yy})^2\right\rangle+\left\langle(\psi_{xx})^2\right\rangle\left\langle(\psi_{yy})^2\right\rangle+\left\langle(\psi_{xy})^4\right\rangle.
\end{align}
Assuming $\DM(x,y)$ to be roughly isotropic, so that $\left\langle(\psi_{xx})^2\right\rangle=\left\langle(\psi_{yy})^2\right\rangle=\left\langle(\psi_{xy})^2\right\rangle$, then
\begin{align}
    \left\langle(\bar\mu_\text{2D})^2\right\rangle= \gamma^2 + \gamma^2 + (\gamma^2)^2 + 3(\gamma^2)^2,
\end{align}
where the factor of 3 on the last term follows from Wick's theorem on the fourth moment. Thus the correlation is
\begin{align}
    r_{\bar\mu_\text{1D},\bar\mu_\text{2D}}=\frac{1}{\sqrt{2+4\gamma^2}}\lesssim\frac{1}{\sqrt{2}},
\end{align}
where the difference from the upper bound is roughly of order the variance of the 1D model magnifications $\ll 1$. A more anisotropic lens has the effect of increasing the correlation, since the last 3 terms in Equation~\ref{eq:autocorr2d} will be effectively smaller. As discussed in Section~\ref{sec:correlation}, other effects that violate the assumptions made in the calculation of $r$ and measurement noise will tend to decrease correlation.

\section{Simulating the fitting procedure}
\label{sec:fittingsim}
To test the fitting procedure, we first generate a spatial gaussian random field over \num{500000} pulses in the Fourier domain $\phi(k)$ as the fourier transform of the `true' DM. The field is $\phi(k)\propto G(k) k^{-3}$, with $G(k)$ drawn from a standard normal distribution, and the proportionality constant set to $\num{1e9}$. We then generate a measured $\DM(x)$ as $\DM(x)=\fourier^{-1}[\phi](x)$, and convert to a simulated measured DM timeseries $\DM(t)$ using $x=v_\text{sim}t$, $v_\text{sim}=\qty{1000}{km.s^{-1}}$, rebin by a factor of 50 to simulate the 50 pulse folding, and add the same level of noise as our measured DM noise.

We also generate a simulated measured magnifications $\mu(x)=1+\frac{\lambda r_e R_F^2}{2\pi}(2\pi i k)^2\fourier^{-1}[\phi](x)$. Two effects are applied to $\mu(x)$: an `intrinsic' smoothing filter taken to be a gaussian filter, and a large-scale multiplicative scintillation/absorption modulation, taken to be a sinusoidal curve. We add same level of noise as our measured magnifications, and convert $\mu(x)$ to $\mu(t)$ using $v_\text{sim}$, and rebin by a factor of 50. Finally, we compute the `background' lightcurve $\mu_\text{bg}(t)$ by smoothing $\mu(t)$.

From there, we can fit $DM(t)$ to $\mu(t)$ using $\mu_\text{bg}(t)$, and the fitting parameters $A$ and $\vscaled$ (the Wiener filter proportionality constant and scaled velocity, respectively) using the procedure described in Section~\ref{sec:implementation}. In 100 simulations, the fitted velocity $\vscaled=\qty{950\pm 50}{km.s^{-1}}$. The bias in our simple 1D simulation can be attributed to the difference between the Wiener-filter, and the `intrinsic' smoothing, here taken as gaussian smoothing.

\end{appendix}

\label{lastpage}

\end{document}